\documentclass[sn-mathphys, Numbered]{sn-jnl}% Math and Physical Sciences Reference Style
\usepackage{caption}
\usepackage{graphicx}%
\usepackage{multirow}%
\usepackage{amsmath,amssymb,amsfonts}%
\usepackage{amsthm}%
\usepackage{mathrsfs}%
\usepackage[title]{appendix}%
\usepackage{xcolor}%
\usepackage{textcomp}%
\usepackage{manyfoot}%
\usepackage{booktabs}%
\usepackage{algorithm}%
\usepackage{algorithmicx}%
\usepackage{algpseudocode}%
\usepackage{listings}%
\usepackage{braket}
\usepackage{amsmath}
\usepackage{mathrsfs}
\usepackage{latexsym}
\usepackage{amssymb}
\usepackage{amsfonts}
\usepackage{bm}

\makeatletter
\newcommand\subsubsubsection{\@startsection{paragraph}{4}{\z@}{-2.5ex\@plus -1ex \@minus -.25ex}{1.25ex \@plus .25ex}{\normalfont\normalsize\bfseries}}
\newcommand\subsubsubsubsection{\@startsection{subparagraph}{5}{\z@}{-2.5ex\@plus -1ex \@minus -.25ex}{1.25ex \@plus .25ex}{\normalfont\normalsize\bfseries}}
\newcommand{\matr}[1]{\mathbf{#1}}
\makeatother

\begin{document}

\title[Improved Financial Forecasting via Quantum Machine Learning]{Improved Financial Forecasting via Quantum Machine Learning}

\author*[1]{Sohum Thakkar}
\author[1]{Skander Kazdaghli}
\author[1,2]{Natansh Mathur}
\author[1,2]{Iordanis Kerenidis}

\author*[3]{André J. Ferreira--Martins}
\author[3]{Samurai Brito}

\affil[1]{QC Ware Corp, Palo Alto, USA and Paris, France}
\affil[2]{IRIF, Université Paris Cité and CNRS, France}
\affil[3]{Itaú Unibanco, São Paulo, Brazil}
\affil[*]{Corresponding authors. Emails: sohum.thakkar@qcware.com, andre.ferreira-martins@itau-unibanco.com.br}

\abstract{Quantum algorithms have the potential to enhance machine learning across a variety of domains and applications. In this work, we show how quantum machine learning can be used to improve financial forecasting. First, we use classical and quantum Determinantal Point Processes to enhance Random Forest models for churn prediction, improving precision by almost 6\%. Second, we design quantum neural network architectures with orthogonal and compound layers for credit risk assessment, which match classical performance with significantly fewer parameters. Our results demonstrate that leveraging quantum ideas can effectively enhance the performance of machine learning, both today as quantum-inspired classical ML solutions, and even more in the future, with the advent of better quantum hardware.}


\maketitle

\section{Introduction}

Quantum computing is a rapidly evolving field that promises to revolutionize various domains, and finance is no exception. There is a variety of computationally hard financial problems for which quantum algorithms can potentially offer advantages \cite{JPM_Survey_QC_Finance, IBM_QC_Finance_Prospects, mckinsey_quantum_computing_industry_use_cases_2021, ProspectsChallengesQuantumFinance}, for example in combinatorial optimization \cite{FinanceRiskNeutralAtom, rebentrost2018quantum}, convex optimization \cite{KerenidisQuantumPortfolioOptimization, QuantumComputataionalFinanceMartingaleAsset}, monte carlo simulations \cite{StochasticOptimalStopping, Suzuki2020QAE, GiurgicaTiron2022lowdepthalgorithms}, and machine learning \cite{JPM_QML_Finance, emmanoulopoulos2022quantum, Alcazar2020ClassicalVsQuantumFinance, nguyen2022bayeseanNNs}.

In this work, we explore the potential of quantum machine learning methods in improving the performance of forecasting in finance, specifically focusing on two use cases within the business of Itaú Unibanco, the largest bank in Latin America. 

In the first use case, we aim to improve the performance of Random Forest methods for churn prediction. We introduce quantum algorithms for Determinantal Point Processes (DPP) sampling \cite{KP2022SubspaceStates}, and develop a method of DPP sampling to enhance Random Forest models. We evaluate our model on the churn dataset using classical DPP sampling algorithms and perform experiments on a scaled-down version of the dataset using quantum algorithms. Our results demonstrate that, in the classical setting, the proposed algorithms outperform the baseline Random Forest in precision, efficiency, and bottom line, and also offer a precise understanding of how quantum computing can impact this kind of problem in the future. The quantum algorithm run on an IBM quantum processor gives similar results as the classical DPP on small batch dimensions but falters as the dimensions grow bigger due to hardware noise.

In the second use case, we aim to explore the performance of neural network models for credit risk assessment by incorporating ideas from quantum compound neural networks \cite{Landman2022quantummethods}. We start by using quantum orthogonal neural networks \cite{Landman2022quantummethods}, which add the property of orthogonality for the trained model weights to avoid redundancy in the learned features \cite{arjovsky2016unitaryNNs}. These orthogonal layers, which can be trained efficiently on a classical computer, are the simplest case of what we call compound neural networks, which explore an exponential space in a structured way. For our use case, we design compound neural network architectures that are appropriate for financial data. We evaluate their performance on a real-world dataset and show that the quantum compound neural network models both have far fewer parameters and achieve better accuracy and generalization than classical fully-connected neural networks. 

This paper is organized as follows: In sections 2-4, we focus on the churn prediction use case and present the DPP-based quantum machine learning methods. In sections 5-7, we present quantum neural network models for risk assessment. Finally, in section 8, we conclude the paper and discuss potential future research directions.

\section{DPP-Enhanced Random Forest Models for Churn Prediction}

\subsection{DPP-Random Forest Model}\label{sec:dpp-rf}

The Random Forest algorithm was introduced in 2001 by Leo Brieman \cite{Breiman2001RF} and has easily become one of the most popular supervised machine learning algorithms in use. It consists of an ensemble of decision trees, each trained on a uniform subsample of rows and columns from the dataset.

In this section, we propose an extension of the Random Forest, called the DPP-Random Forest (DPP-RF), which utilizes Determinantal Point Processes (DPPs) instead of uniform sampling to subsample rows and columns for individual decision trees. In the original RF algorithm, subsampling makes the model more robust to variance in the training data; however, the use of uniform sampling runs the risk of improperly representing the dataset and missing under-sampled areas. DPP sampling better preserves the diversity of the dataset, and corrects for sampling bias \cite{Kulesza2012DPPforML}. We first introduce the theory and techniques of DPP sampling, then present the algorithm.

\subsection{Determinantal Point Processes}
We will now introduce the Determinantal Point Process (DPP), which lies at the core of the methodology behind our solution to the churn problem. DPPs are a class of probabilistic models that can be used to sample diverse subsets of items from a larger set. They were first formalized by Macchi in 1975 as a way to model fermions in quantum mechanics \cite{macchi_1975}. More recently, these models are showing increasing promise in the context of machine learning \cite{Kulesza2012DPPforML}, where they can be used for a variety of tasks, such as building unbiased estimators for linear regression \cite{derezinski2021determinantal}, performing monte carlo estimation \cite{bardenet2020monte}, and promoting diversity in model outputs \cite{elfeki2019gdpp}.

\subsubsection{Definitions}
A point process $P$ on a set $Y$ is a probability measure over the subsets of $Y$. Sampling from a point process on $Y$ will produce some subset $S \subseteq Y$ with probability $P(S)$. A repulsive point process is a point process in which points that are more similar to each other are less likely to be selected together.

A determinantal point process (DPP) is a particular case of a repulsive point process, in which the selection probability of a subset of items $T \subseteq Y$ is given by a determinant. Given a real, symmetric $n \times n$ matrix $K$ indexed by the elements of $Y$: 
\[ P\{T \subseteq S \} = \det(K_{T,T}) \ , \]
where $K_{T,T}$ denotes the $|T | \times |T |$ submatrix indexed by the set $T$ and $n$ is the cardinality of $Y$. In other words, the marginal distribution $P\{T \subseteq S\}$ is defined by the subdeterminants of K.

The above is the most general definition, but in machine learning, we typically focus on a slightly more restrictive class of DPPs called $L$-ensembles. In $L$-ensembles, the whole distribution, not just the marginals, is given by the subdeterminant of a real, symmetric  $n \times n$ matrix $L$. \[P\{S\} \propto \det(L_{S,S})\ . \]
Just like $K$, $L$ is indexed by the elements of $Y$. Because of some convenient properties of the determinant \cite{Kulesza2012DPPforML}, we can explicitly write down the distribution of an $L$-ensemble:
\[P\{S\} = \frac{\det(L_{S,S})}{\det(L+I)} \ . \]
In machine learning literature, DPPs are typically defined over a set of points $\matr{X}$, with each item $\matr{x}_i$ a row in the data matrix $\matr{X}$. If we preprocess $\matr{X}$ such that its columns are orthonormal and choose $\matr{L}$ to be the inner-product similarity matrix, i.e. $L = \matr{X}\matr{X}^T$,  then the distribution becomes even simpler to write down. Instead of explicitly computing the $\matr{L}$ matrix, we can write the distribution in terms of the data matrix $\matr{X}$ itself, courtesy of the Cauchy-Binet formula,

\begin{equation} \label{eq:dpp_dist_xxt}
P\{S\} = \frac{\det(\matr{X}_S)^2}{\det(\matr{X}\matr{X}^T + \matr{I})} \ .
\end{equation}

Moreover, the distribution will almost surely produce samples with size $d$, the rank of the orthogonalized data matrix $X$. This kind of DPP is denoted $d$-DPP. We will focus here on an application of sampling from a $d$-DPP from a data matrix X.

\subsubsection{Unbiased Least-Squares Regression}
One unique feature of the DPP compared to i.i.d sampling techniques is that it can lead to provably unbiased estimators for least-squares linear regression \cite{Derezinski2018LinearRegression} \cite{Derezinski2018LeveragedForLinearRegression}. Given an $n\times d$ data matrix $\matr{X}$ and a target vector $\matr{y} \in \mathbb{R}^n$, where $n \gg d$, we wish to approximate the least squares solution $\matr{w}^{\ast} = \mathrm{argmin}_\matr{w} ||\matr{X}\matr{w} - \matr{y}||$. $\matr{w}^{\ast}$ represents the best-fit parameters to a linear model to predict $\matr{y}$.

Surprisingly, if we sample $d$ points $S$ from DPP($\matr{X}\matr{X}^T$) and solve the reduced system of equations $\matr{y}_S = \matr{X}_S\matr{w}$, we get an unbiased estimate of $\matr{w}^\ast$. Formally, if $S \sim d\text{-DPP}_\matr{L}(\matr{X}\matr{X}^{T})$, 
\begin{equation}
    \mathop{\mathbb{E}}[\matr{X}_S^{-1} \matr{y}_S] = \mathrm{argmin}_\matr{w} ||\matr{X}\matr{w} - \matr{y}|| = \matr{w}^{\ast} \ . 
\end{equation}

This allows us to create an "ensemble" of unbiased linear regressors, each trained on a DPP sample. In some regard, this was the inspiration for trying an ensemble of \emph{decision trees} trained on DPP samples, as detailed in Sec. \ref{sec:dpp-rf}.

\subsubsection{Algorithms for Sampling}
There are several efficient algorithms for sampling from DPPs and computing their properties.
The naive sampling method --- calculating all subdeterminants and performing $l2$ sampling --- takes exponential time. The first major leap in making DPP sampling feasible on today's computers was the "spectral method" \cite{Kulesza2011kDPP, Hough2006DPPOriginalAlgorithm}. This algorithm performs an eigendecomposition of the kernel matrix before applying a projection-based iterative sampling approach. Thus, the first sample takes $O(nd^2)$ time, and subsequent samples take $O(d^3)$.

Monte Carlo methods have been proposed to approximate the DPP distribution \cite{Anari2016MonteCarloKDPP, Li2016MonteCarloDPP}, though they are not exact, and are often still prohibitively slow with a runtime of $O(n \text{poly}(d))$ per sample.

In a counter-intuitive result, \cite{NEURIPS2019_VFX} and \cite{NEURIPS2020_ALPHA} proposed methods that avoid performing the full DPP sampling procedure on large parts of the basis set. This approach resulted in a significant reduction in runtime, making DPPs more practical for mid-to-large-scale datasets. These techniques allow exact sampling of subsequent $d$-DPP samples in $O(\text{poly}(d))$, independent of the size of the full basis set $n$. Many of these algorithms are implemented in the open-source DPPy library\cite{Gautier2019DPPy}, which we used in the experiments in this paper. 

Recent work has shown that quantum computers are in principle able to sample from DPPs in even lower complexity in some cases. We describe this quantum algorithm in Sec. \ref{sec:qdpp}. This and several other algorithms which arise from the techniques introduced in  \cite{KP2022SubspaceStates} are a budding area of research in the quantum computing space, and will hopefully inspire more applications like the one we describe in this paper. For example, in \cite{kazdaghli2023improved}, DPPs and deterministic DPPs were used to improve the methods for the imputation of clinical data.

\subsection{DPP-RF Algorithm Outline}
In principle, the DPP-RF uses DPP sampling on the whole dataset to select diverse subsets of data on which to train decision trees. However, sampling from a DPP on large datasets (like the entire churn dataset of 174,000 points) can take copious time, especially when using current open-source implementations of DPP sampling. To be able to test these techniques quickly, a novel sampling procedure was developed which preserves many of the benefits of DPPs, but does not require sampling from the full dataset. The procedure can be summarized as follows:

\begin{enumerate}
    \item Divide the training set uniformly into smaller batches;
    \item Sample $S_1 \sim d$-DPP$(\matr{X}_{batch}\matr{X}_{batch}^T)$ data points from every batch;
    \item Sample $S_2 \sim d$-DPP$(\matr{X}_{S_1}^T \matr{X}_{S_1})$ features;
    \item Train a first group $G_1$ of $N_1$ decision trees on these small patches of data;
    
    % \begin{figure}[H]
    %     \centering
    %     \includegraphics[width=.8\linewidth]{pics/step1.png}
    %     \caption{Steps 1 to 4 of the DPP-Random Forest algorithm.}
    %     \label{fig:algo1}
    % \end{figure}

    \item Aggregate the patches of data resulting from step 2 to create a larger dataset $\matr{X}_{agg}$;
    \item Repeat for $N_2$ times: sample $S_3 \sim d$-DPP$(\matr{X}_{agg}^T \matr{X}_{agg})$ features to create a long matrix;
    \item Train a second group $G_2$ of $N_2$ decision trees on these new datasets;
    
    % \begin{figure}[H]
    %     \centering
    %     \includegraphics[width=.8\linewidth]{pics/step2.png}
    %     \caption{Steps 5 to 7 of the DPP-Random Forest algorithm.}
    %     \label{fig:algo2}
    % \end{figure}
    
    \item Combine $G_1$ and $G_2$ by aggregating them to make predictions (similar to the classical Random Forest algorithm).

\begin{figure}[H]
    \centering
    \includegraphics[width=\linewidth]{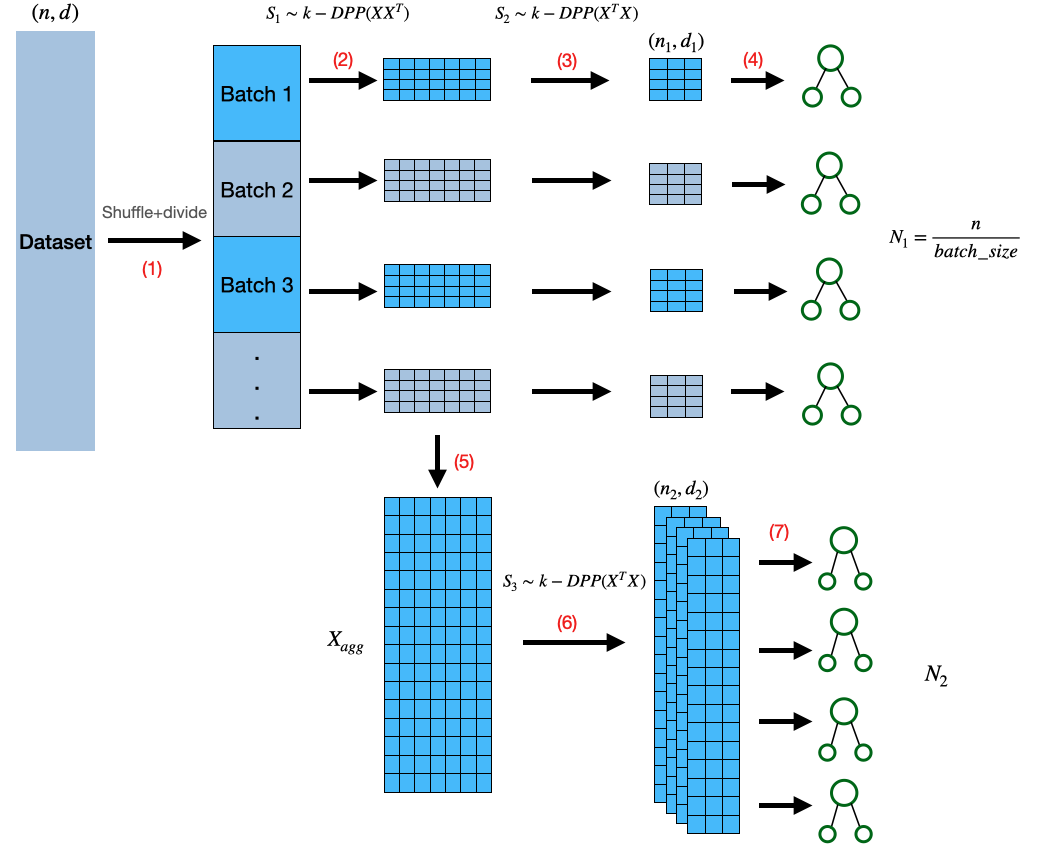}
    \caption{Steps 1 to 7 of the DPP-Random Forest algorithm.}
    \label{fig:dpp-algo}
\end{figure}

\end{enumerate}

\section{Classical DPP-RF Results}\label{sec:results}
The DPP-RF algorithm was designed for the purpose of predicting customer churn in the bank. In this section, we define this use case and present the results. In addition, we benchmark our proposed DPP-RF method by constructing models on public tabular classification tasks.

\subsection{Use Case Introduction: Churn Prediction}
Churn, defined as a customer withdrawing more than a certain amount of money in a single month, is a significant concern for retail banks. Our objective is to predict which customers are most likely to churn in the next three months using customer data from the previous six months.

The primary dataset used in this study consists of 304,000 datapoints, with 153 features for each datapoint. Each datapoint represents a banking customer at a particular month in time, with the features representing various aspects of their activity over the previous six months. The target variable is a binary flag indicating whether or not the customer churned in at least one of the following three months. The data was anonymized and standardized before being split into training and test sets based on time period, with 130,000 datapoints being set aside as the test set and 174,000 datapoints used for training. The data was split in a way that did not produce any significant covariate shift between the train and test sets.

With the end goal of preventing churn, the model works by flagging customers with the highest risk of potential churn. For these flagged customers, the bank can deploy a representative to intervene and better understand their needs. However, resource limitations make it necessary to flag a relatively small number of customers with high confidence. The focus of this exploration was to reduce false positives in the flagged customers to increase the efficiency of bank interventions. In terms of the precision-recall trade-off, our model should be tuned to provide the highest possible precision for low recall values. Despite this simplification to a classification problem, the primary business KPI is the amount of withdrawal money correctly captured by the model, as discussed more in Sec. \ref{sec:withdrawals_captured}.

This use case already had a solution in production: a Random Forest classifier \cite{Breiman2001RF}, whose performance was used as a benchmark. The model in production already captured a significant amount of churn, but there was clear room for improvement in the amount of withdrawals captured (see Fig. \ref{fig:bench1}). Moreover, given the large number of customers in the dataset and the relative homogeneity of the population of interest, there existed an opportunity to employ techniques that explicitly try to explore diversity in the data.

We focused on three key performance indicators (KPIs): the precision-recall curve, the training time and the bottom line. 

\subsection{Precision-recall}\label{sec:prec-recall}
\begin{figure}[H]
    \centering
    \includegraphics[width=.8\linewidth]{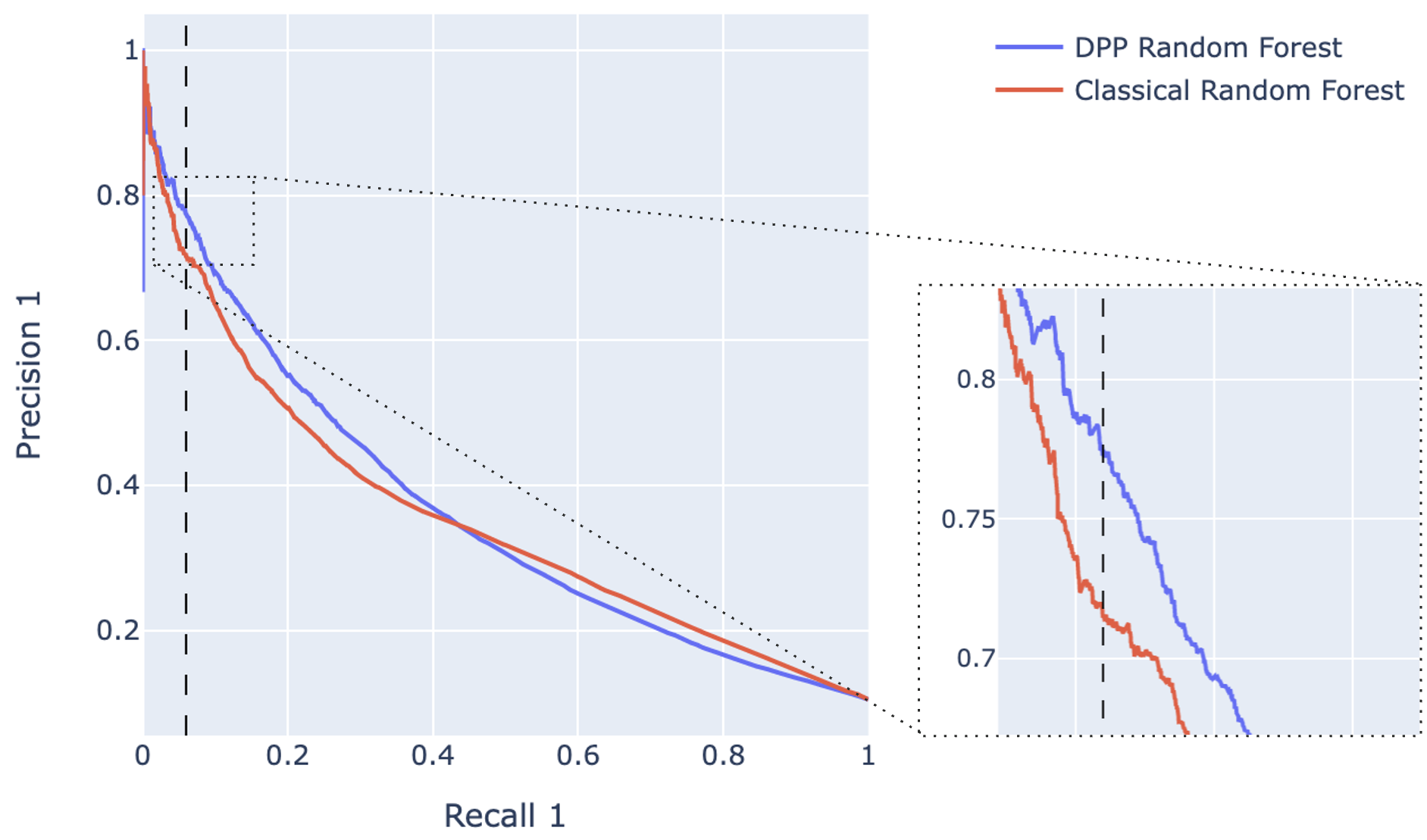}
    \caption{Precision-recall curve for the test set. Using DPP with the Random Forest algorithm shows an improvement of $5.9\%$.}
    \label{fig:pr}
\end{figure}

To evaluate the performance of our proposed method, we optimized hyperparameters and measured the precision for a low fixed recall (6\% in this case). As seen in Figure \ref{fig:pr}, our method showed an improvement in precision from $71.6\%$ for the benchmark model to $77.5\%$ with the new model. Our method also provided similar improvements in precision for the relevant range of small recall.

\subsection{Training Time}

The DPP-RF model has a longer training time compared to the traditional random forest on a classical computer: it took 54 minutes to train the model with the best hyperparameters using, compared to 311 seconds for the benchmark model. The models were trained on a computer with an Intel\copyright \ Core\texttrademark \ i5-8350U CPU running at 1.70 GHz, 24 GB of RAM and Windows 10 version 21H2, compilation 19044.2604.

The computational bottleneck in this algorithm is the DPP sampling. Instead of simulating quantum DPP circuits (which is infeasible for large datasets), we used a classical SVD-based sampling algorithm \cite{Hough2006DPPOriginalAlgorithm} implemented in the dppy library \cite{Gautier2019DPPy}. We believe that improved classical sampling techniques \cite{NEURIPS2020_ALPHA} and future quantum techniques (Sec. \ref{sec:qdpp}) can reduce the runtime dramatically.

Hyperparameters were selected using a grid search with 5-fold cross-validation over the typical RF parameters n\_estimators, max\_depth, min\_samples\_leaf, min\_samples\_split, max\_features, max\_samples, and the DPP-RF-specific parameter batch\_size. The training time of a DPP-RF model depends heavily on this batch\_size parameter, which is the size of the batches from which we take DPP samples. Choosing a batch size higher than 1000 can increase runtime dramatically. Thus, in our hyperparameter search, we limited the batch size to less than 1000.

Within the bank, the churn model is retrained just once every few months, so the training time was not prohibitive. However, faster sampling algorithms still serve to increase the range of feasible hyperparameters (especially the batch size). 

\subsection{Bottom Line --- Withdrawals Captured}\label{sec:withdrawals_captured}
\begin{figure}[H]
    \centering
    \includegraphics[width=.85\linewidth]{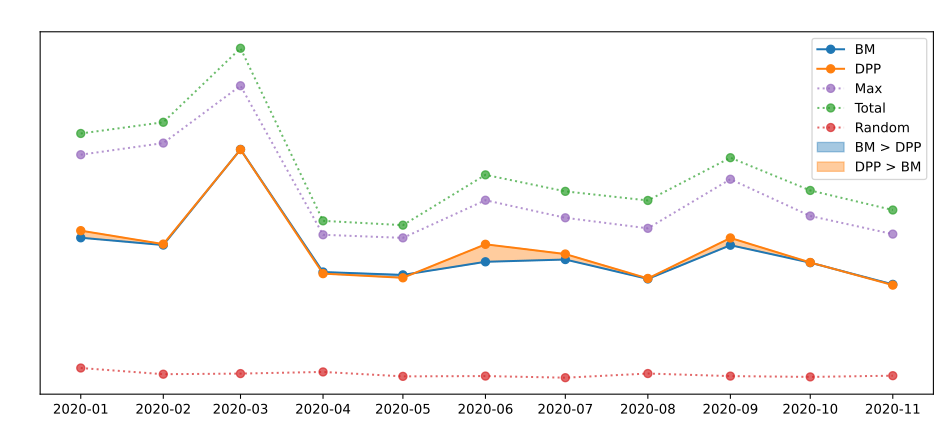}
    \caption{Classical benchmark (BM) vs DPP-RF solution: money withdrawn per month by the flagged 500 customers, comparing the benchmark model (blue line) to the DPP-RF one (orange line). On the y axis, we have monetary values (not shown). The green line represents the total amount of money withdrawn by all customers in each month. The purple line is the sum of the 500 largest withdrawals, which is the maximum value that the model could capture. The red line represents the withdrawals captured by randomly flagging 500 observations. The y-axis units are omitted for confidentiality.}
    \label{fig:bench1}
\end{figure}

\begin{figure}[H]
    \centering
    \includegraphics[width=.9\linewidth]{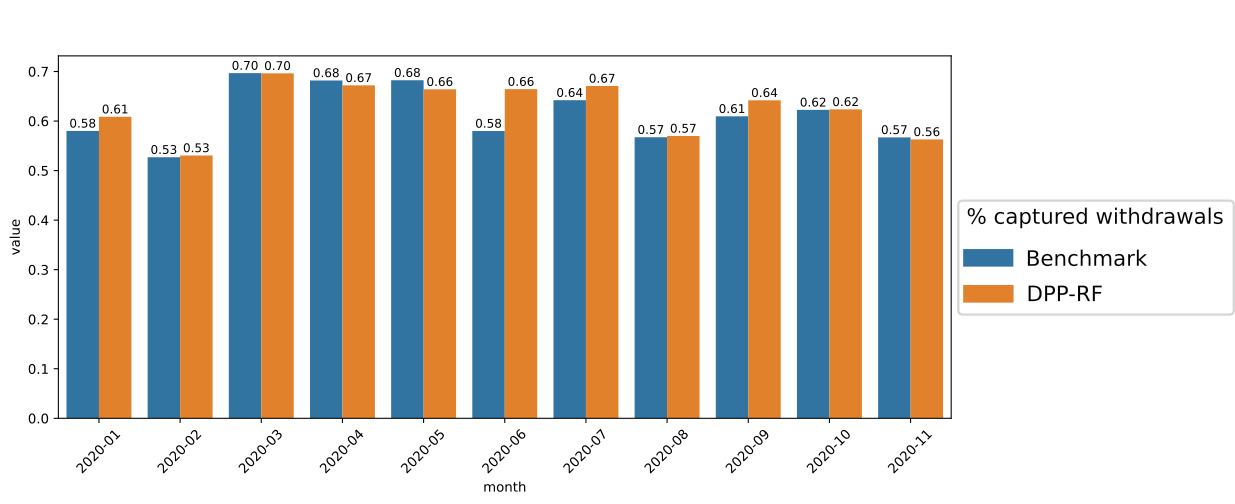}
    \caption{Classical benchmark vs DPP-RF solution - percentage of total withdrawals captured per month, that is, relative to the green line in Fig. \ref{fig:bench1}. On average over the 11 test months, the BM model captures 61.42\% of the total, whilst the DPP-RF model captures 62.77\% --- an improvement of 1.35\%.}
    \label{fig:bench2}
\end{figure}

\begin{figure}[H]
    \centering
    \includegraphics[width=.9\linewidth]{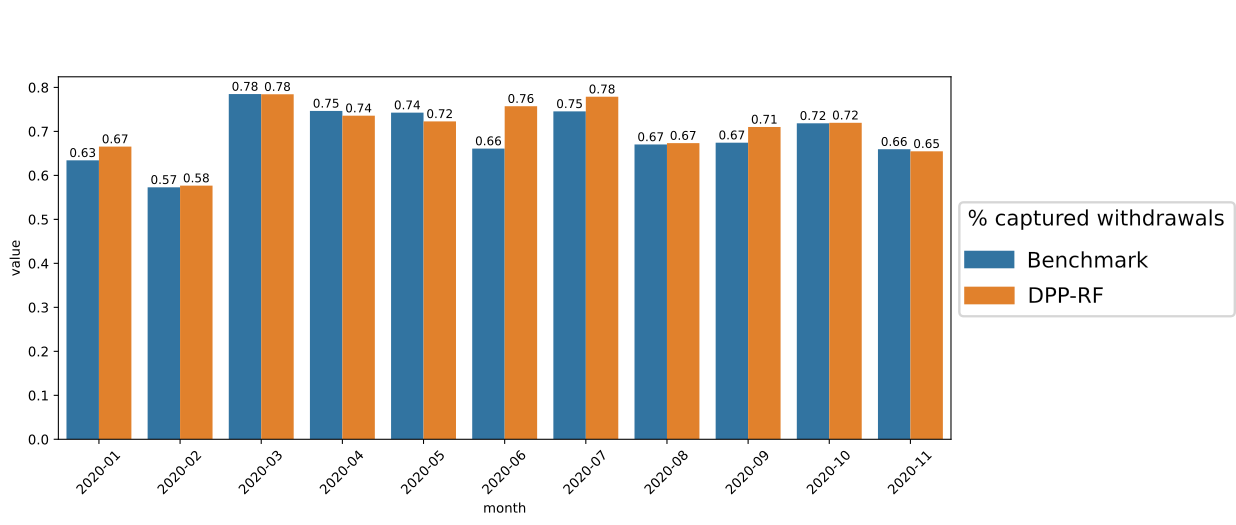}
    \caption{Classical benchmark vs DPP-RF solution - the percentage of maximum money possible to be captured (given $\text{n\_flags} = 500$ customers flagged every month), that is, relative to the purple line in Fig. \ref{fig:bench1}. On average over the 11 test months, the BM model captures 69.18\% of the total, whilst the DPP-RF model captures 70.72\% --- an improvement of 1.54\%.}
    \label{fig:bench3}
\end{figure}

From a business perspective, the most direct indicator of the success of the model is the amount of assets under management (AUM) that can be salvaged via interventions. Thus, we evaluated the amount of money withdrawn every month by the 500 customers flagged by the model, i.e., the 500 customers which had the highest predicted probability of churning in one of the following 3 months. As seen in figures \ref{fig:bench1}, \ref{fig:bench2} and \ref{fig:bench3}, our model showed substantial overall improvements. The true financial impact of these predictions is dependent on the success of the interventions as well as the bank's profit-per-dollar-AUM.

\subsection{Summary of Results}

The proposed DPP Random Forest model provides significant improvements in precision and bottom line, while taking significantly longer to train. The results are summarized in the following table.

\begin{table}[h]
\caption{Summarized comparison between models.}\label{table:summary_comparison}
\begin{tabular}{@{}p{0.45\textwidth}p{0.21\textwidth}p{0.21\textwidth}@{}}
\toprule
Metric & Benchmark model & Proposed model \\
\midrule
Precision  &71.6\%&77.5\% \\
\% total withdrawals captured &61.42\%&62.77\%\\
\% maximum possible withdrawals captured &69.18\%&70.72\%\\
Train time &311s& 54 min \\
\botrule
\end{tabular}
\end{table}

\subsection{Further Benchmarks}
\label{sec:further_benchmarks_rf}

We further benchmarked our model on various classification datasets from OpenML. All except one (madelon) of these datasets were used in \cite{grinsztajn2022why} and preprocessed accordingly. They were chosen to be representative of a wide variety of classification tasks. Each dataset was split into train, validation, and test sets. For each model, 400 sets of hyperparameters were randomly chosen and evaluated on the validation set. Both models used the same hyperparameter space, except for the addition of the batch\_size parameter for the DPP-RF. The hyperparameters which gave the best results on the validation set were evaluated on the test set, and the results are reported below. Models were evaluated with the ROC-AUC metric\footnote{The area under the receiver operating characteristic curve (ROC-AUC) is a common metric for two-class classification tasks, and evaluates the ability of the model to produce a proper ranking of datapoints by likelihood of being class 1.}.

\begin{table}[h]
\caption{Comparison of DPP Random Forest and Random Forest models for different datasets. The results are reported via the ROC-AUC metric.}\label{table:DPP_RF_comparison}
\begin{tabular}{@{}p{0.25\textwidth}p{0.25\textwidth}p{0.25\textwidth}@{}}
\toprule
Dataset & Random Forest & DPP Random Forest \\
\midrule
madelon & 0.916 & \textbf{0.941} \\
credit-default & 0.856 & 0.856 \\
house-pricing & \textbf{0.948} & 0.939 \\
jannis & 0.866 & \textbf{0.870} \\
eye movements & 0.704 & \textbf{0.710} \\
bank-marketing & {0.881} & 0.881 \\
wine & 0.901 & \textbf{0.906} \\
california & 0.962 & \textbf{0.963} \\
\botrule
\end{tabular}
\end{table}

\section{Quantum DPP-RF}\label{sec:qdpp}

\subsection{Quantum Circuits for Determinantal Point Processes}\label{sec:dpp_circuit}

Classical DPP sampling algorithms have improved significantly since their inception, but it still remains infeasible to sample from large datasets like ours. Recent work by Kerenidis and Prakash \cite{KP2022SubspaceStates} has shown that a quantum computer can more natively perform DPP sampling, achieving a gate complexity of $O(nd)$ and a circuit depth of $O(d\log(n))$ for an orthogonal matrix of size $n \times d$. The classical time complexity for sampling is $O(d^3)$ \cite{Hough2006DPPOriginalAlgorithm}. Note that when $n$ is very large, then one can reduce the number of rows to $O(d^2)$ before performing the sampling \cite{NEURIPS2020_ALPHA}. 
%If the matrix is not orthogonal, the Quantum DPP algorithm still produces DPP samples but fails transparently with some probability. It is a future direction to bound the failure probability, which in practice is reasonable.

For a thorough review of the quantum methods and circuits, we refer the reader to \cite{KP2022SubspaceStates}. The circuit is described in brief below.

Given an orthogonal matrix $\matr{X} = (\matr{x}^1, \matr{x}^2, \dots, \matr{x}^n) \in \mathbb{R}^{n\times d}$, the quantum DPP circuit applied on $\matr{X}$ performs the following operation:

\begin{equation}
    \mathcal{D}(\matr{X})|0^n \rangle = \sum_{\substack{|S|=d \\
                                                S \in \{0,1\}^n}} 
                                        \det(\matr{X}_S)| 1_S \rangle \ ,
\end{equation}

\noindent where $\matr{X}_S$ is the $\mathbb{R}^{d\times d}$ submatrix obtained after sampling the rows of $\matr{X}$ indexed by $S$; $1_S$ is the characteristic vector of $S$ (with 1's in the positions indexed by the elements of S) and $\mathcal{D}(\matr{X})$ represents the quantum $d$-DPP circuit, as detailed below.

\begin{figure}
    \centering
    \begin{minipage}{0.45\textwidth}
        \centering
        \includegraphics[width=\textwidth]
        {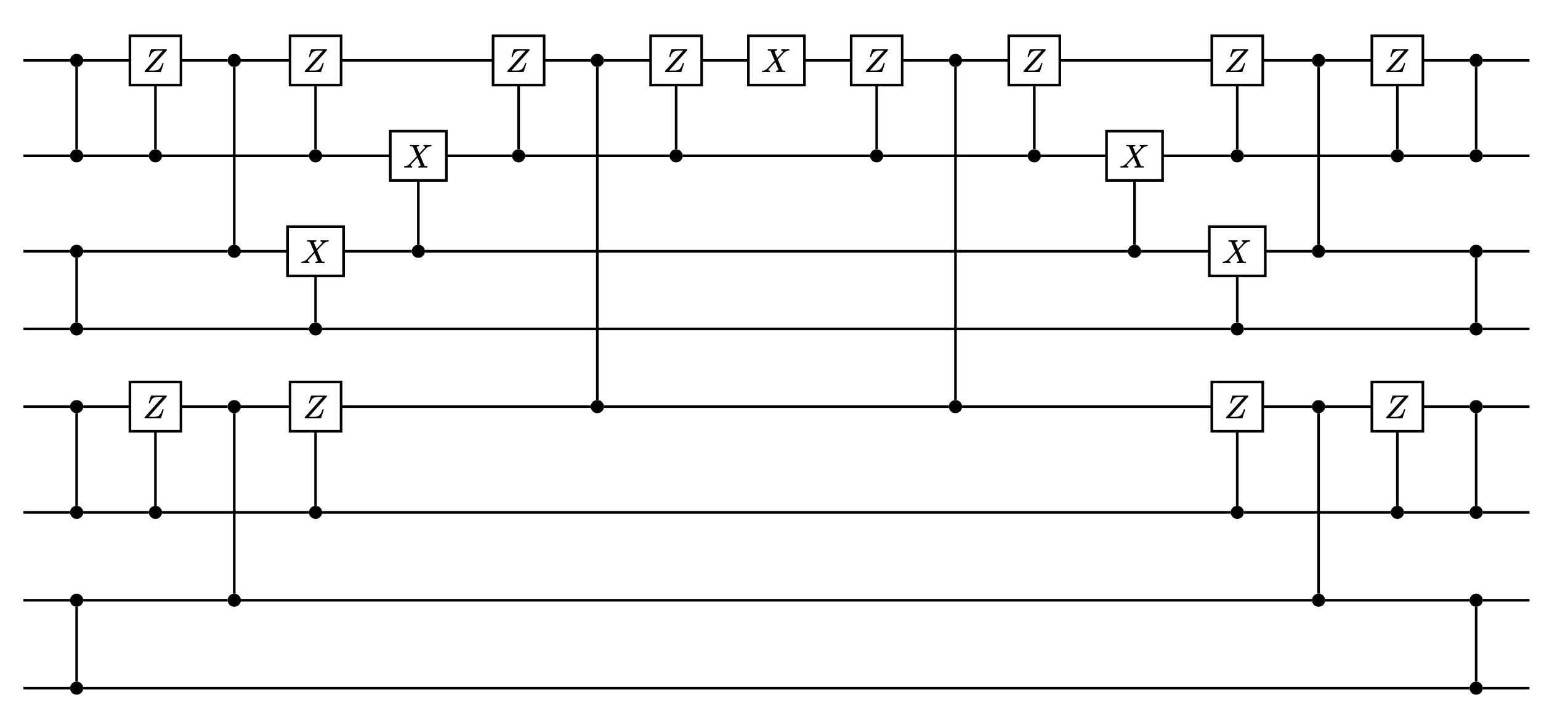}
        \caption{Clifford Loader circuit $\mathcal{C}(x)$ for $x \in \mathbb{R}^8$.}
        \label{fig:parallel_clifford}
    \end{minipage}
    \hspace{0.1cm}
    \begin{minipage}{0.48\textwidth}
        \centering
        \includegraphics[width=0.75\textwidth]{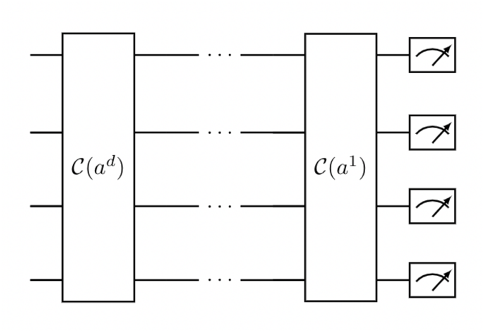}
        \caption{DPP circuit as a series of Clifford Loaders.}
        \label{fig:qdpp_circuit}
    \end{minipage}
\end{figure}

Thus, the probability of sampling $S$, i.e., of measuring $|1_S \rangle$, is: $Pr(S) = \det(\matr{X}_S)^2 = \det(L_{S,S})$, where $L=\matr{X}\matr{X}^T$. This draws the link between the quantum determinantal sampling circuit and the classical $d$-DPP model as seen in eq. \ref{eq:dpp_dist_xxt}.

To construct the quantum $d$-DPP circuit, we need to first introduce a circuit known as a \textit{Clifford loader}, which performs the following operation:

\begin{equation}
    \mathcal{C}(\matr{x}) = \sum_{i=1}^n x_i Z^{i-1} X I^{n-i}, \quad \text{for}  \quad \matr{x} \in \mathbb{R}^n \ .
\end{equation}

The Clifford loader was shown to have a log-depth circuit in \cite{KP2022SubspaceStates}, and is shown for $n=8$ in fig \ref{fig:parallel_clifford}, in which the gates represented by vertical lines are RBS gates --- parameterized, hamming weight preserving two-qubit gates.

The full quantum $d$-DPP circuit is a series of $d$ Clifford loaders, one for each orthogonal column of $\matr{X}$:

\begin{equation}
    \mathcal{D}(\matr{X}) = \mathcal{C}(\matr{x}^1) \mathcal{C}(\matr{x}^2) \dots \mathcal{C}(\matr{x}^d) \ .
\end{equation}

An example of a $d$-DPP circuit as a series of Cliffords for $n=4$ is shown in fig. \ref{fig:qdpp_circuit}.

\subsection{Hardware Experiment Results}

As a hardware experiment, we aimed to implement a simplified version of our algorithm on a quantum processor. We chose to use the "ibmq\_guadelupe" 16-qubit chip, which is only capable of running small quantum DPP circuits for matrices of certain dimensions, such as $(4,2), (5,2), (5,3), (6,2), (8,2)$. As a result, we had to reduce the size of our problem.

\begin{figure}[H]
    \centering
    \includegraphics[width=.8\linewidth]{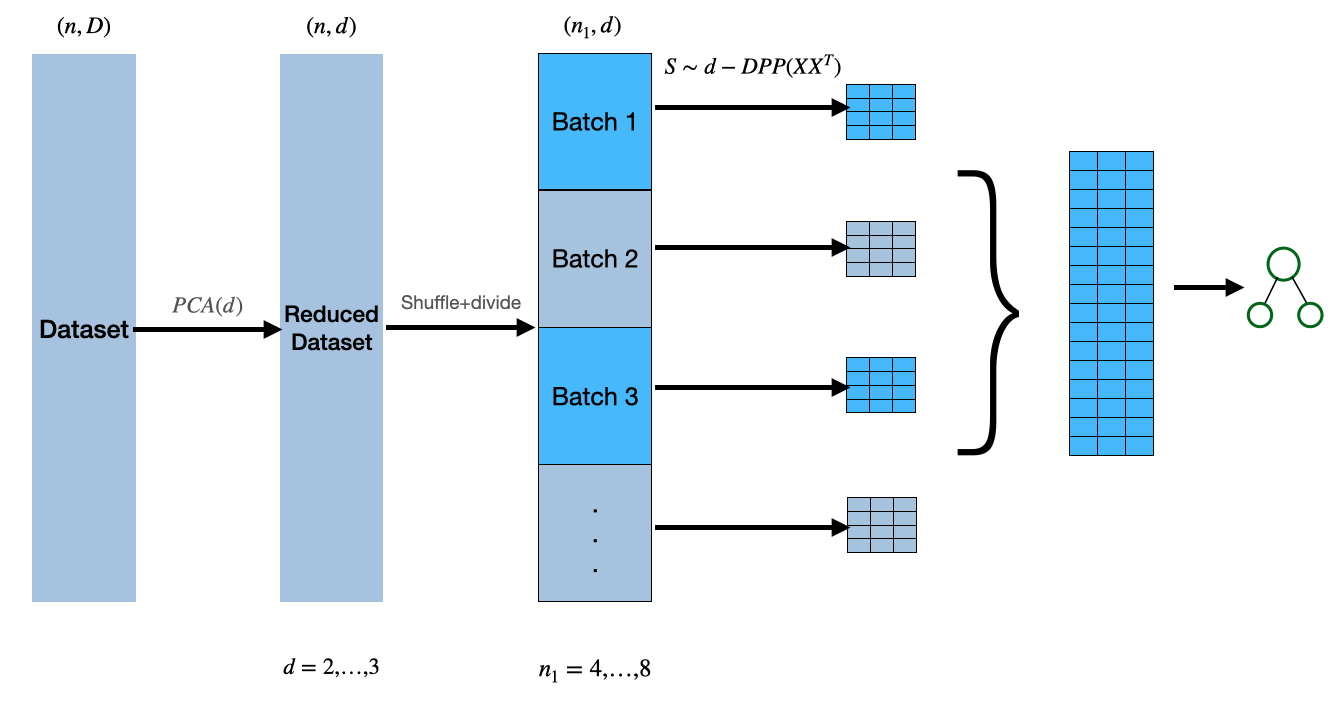}
    \caption{Quantum hardware-ready procedure for DPP sampling.}
    \label{fig:hrd}
\end{figure}

To accomplish this, we defined reduced train/test sets: a train set of $\sim$1000 points from 03/2019 and a test set of $\sim$10000 points from 04/2019. The quantum hardware-ready simplified algorithm is outlined in Figure \ref{fig:hrd}. It includes the following steps:

\begin{enumerate}
\item Applying PCA to reduce the number of columns from 153 to $d=2,3$;
\item Dividing the dataset into batches of $n=4,5,6,8$ points;
\item Sampling $S \sim d$-DPP$(\matr{X}_{batch}\matr{X}_{batch}^T)$ rows from each batch, resulting in small $d\times d$ patches of data;
\item Aggregating these patches to form a larger dataset, then training one decision tree on this dataset.
\end{enumerate}

We repeated this process for a number of trees and estimated the F1\footnote{We chose the F1 score as the evaluation metric for two reasons. Firstly, a single decision tree, unlike the random forest, does not provide an estimate of the likelihood, which is required for the computation of ROC-AUC as we had used before. Secondly, we had an imbalanced dataset and thus needed a metric that balanced precision and recall. The F1 score is particularly effective when used in these scenarios.} score for every tree. We then compared the results for different sampling methods: uniform sampling, quantum DPP sampling using a simulator, and quantum DPP sampling using a quantum processor.

\begin{figure}[H]
    \centering
    \begin{minipage}{0.45\textwidth}
        \centering
        \includegraphics[width=\textwidth]
        {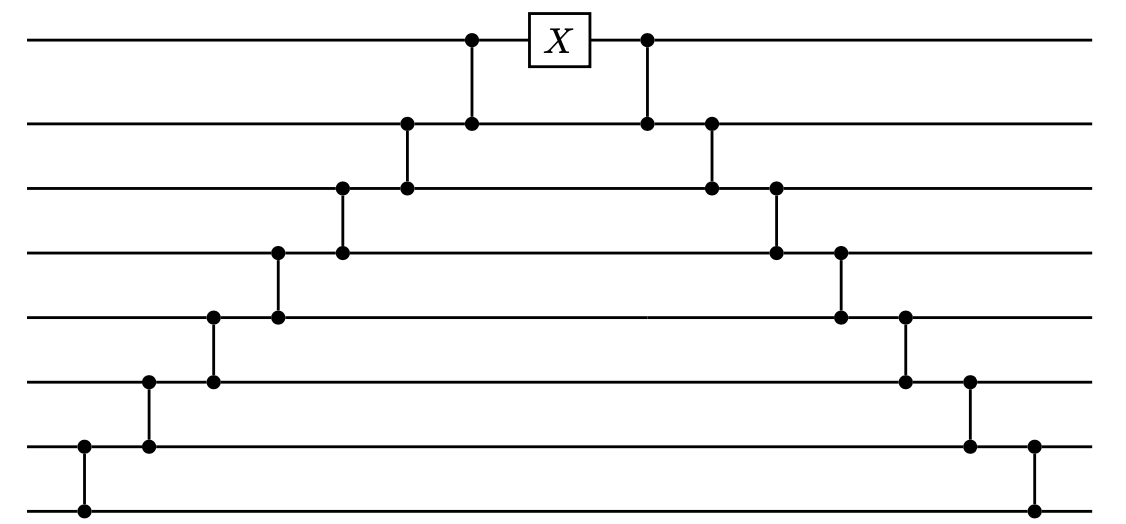}
        \caption{Diagonal Clifford loader.}
        \label{fig:diagonal_clifford}
    \end{minipage}
    \hspace{0.1cm}
    \begin{minipage}{0.45\textwidth}
        \centering
        \includegraphics[width=0.75\textwidth]{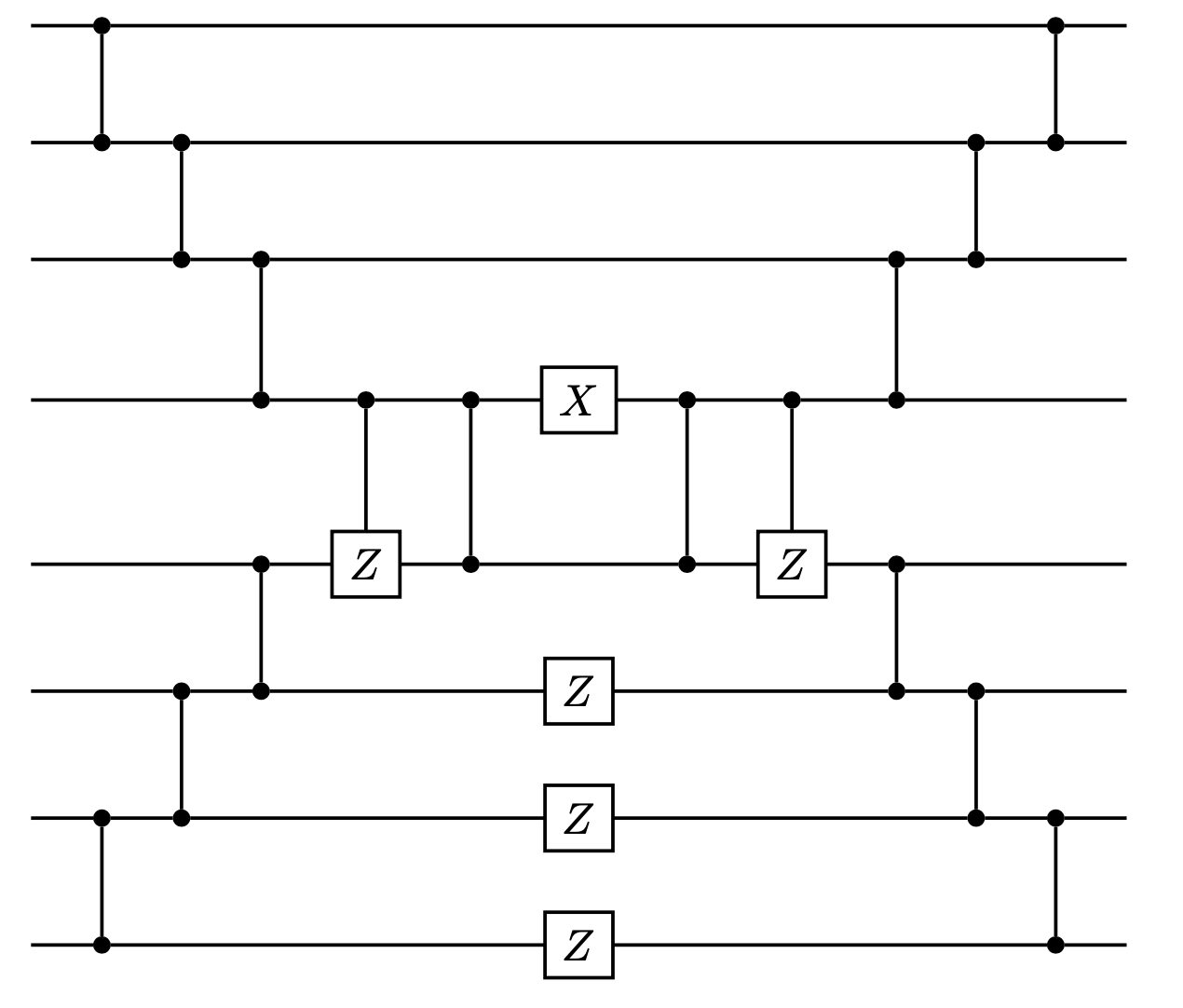}
        \caption{Semi-diagonal Clifford loader.}
        \label{fig:semidiagonal_clifford}
    \end{minipage}
\end{figure}

\begin{figure}[H]
    \centering
    \includegraphics[width=\linewidth]{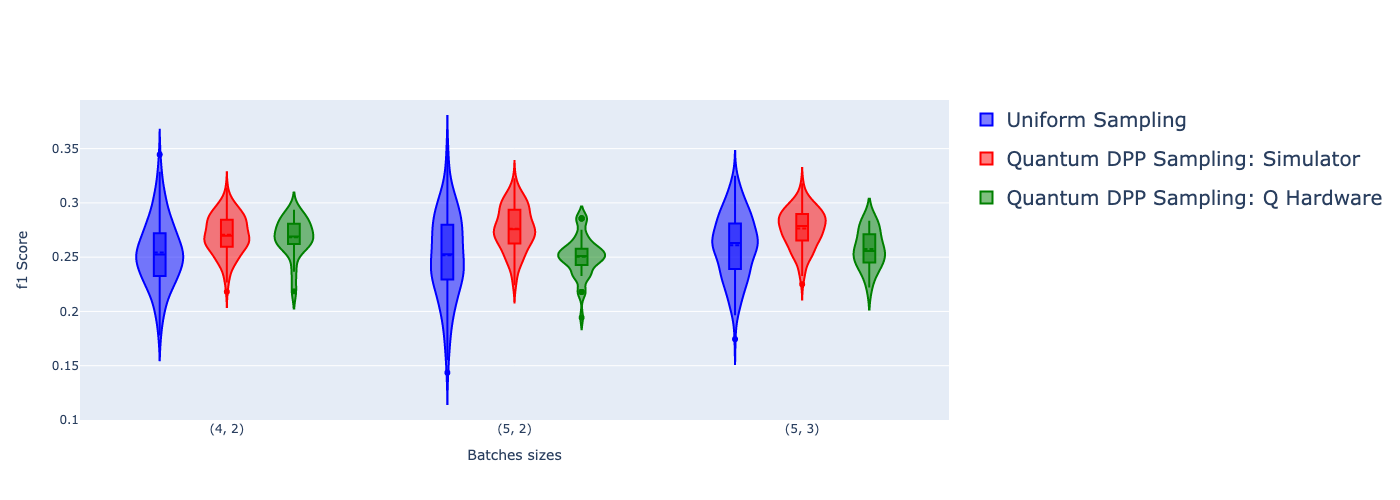}
    \caption{Decision trees performance using Quantum DPP sampling with diagonal Clifford loaders.}
    \label{fig:f1diag}
\end{figure}

\begin{figure}[H]
    \centering
    \includegraphics[width=\linewidth]{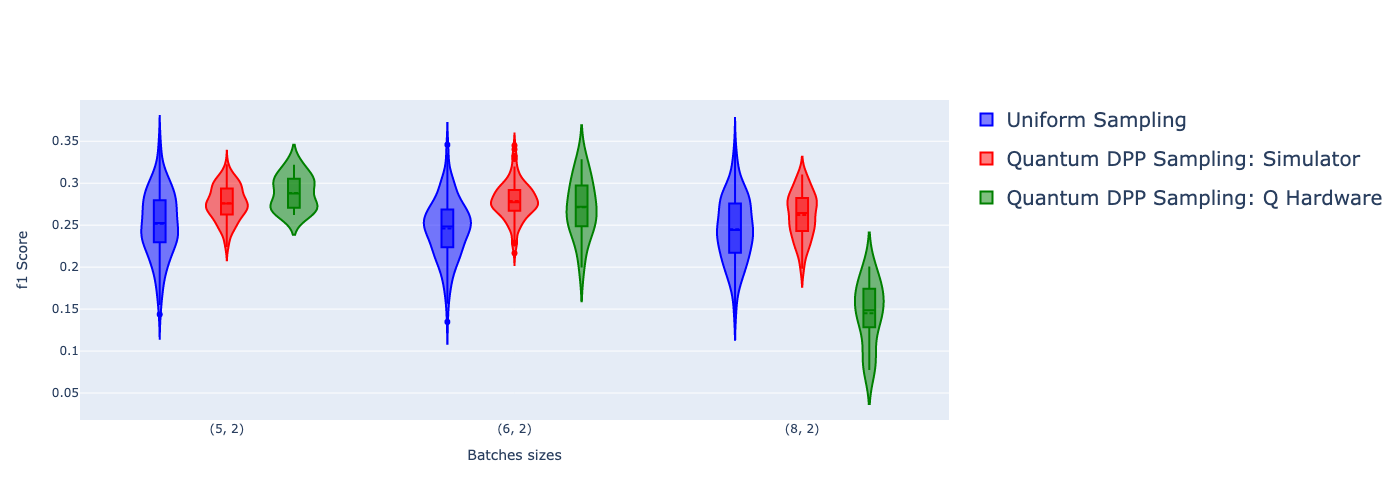}
    \caption{Decision trees performance using Quantum DPP sampling with semi-diagonal Clifford loaders.}
    \label{fig:f1semi}
\end{figure}

The IBM quantum computer only allows using RBS gates on adjacent qubits, so we cannot use the circuit described in section \ref{sec:dpp_circuit}. Instead, we use two different Clifford loader architectures which only use adjacent-qubit connectivity, visualized in fig \ref{fig:diagonal_clifford}. The diagonal Clifford loader is explained in \cite{KP2022SubspaceStates}, and the semi-diagonal loader is a modification that halves the circuit depth. As an error mitigation measure, we disregarded all results that did not have the expected hamming weight ($d$). The results are shown in the violin plots in Figures \ref{fig:f1diag} and \ref{fig:f1semi}.

The results indicate that for small matrix dimensions --- up to (6,2) --- the IBM quantum processor gives results similar to the ones achieved with the simulator. However, as the dimensions grow bigger, the samples from the quantum DPP circuits lead to worse classifier performance. This highlights the limitations of the available quantum hardware, which is prone to errors.

\section{Quantum Neural Networks for credit risk assessment}

\subsection{Quantum Neural Networks with Orthogonal and Compound Layers}

In recent years, variational/parameterized quantum circuits \cite{pqc} have become very prominent as NISQ-friendly QML techniques. When applied to classification problems, they are commonly known as Variational Quantum Classifiers (VQC) \cite{ibm}. The quantum circuits associated with VQCs may be schematically thought of as composed of three layers: the
\textit{feature map} $\mathcal{U}_{\Phi(\vec{x})}$, which encodes classical data $\vec{x}$ into quantum states; the
\textit{variational layer} $W(\bm{\theta})$, which is the part of the circuit parameterized by a set of parameters $\bm{\theta}$ which are learned in the training process; and finally, the \textit{measurement layer}, which measures the quantum registers and produces classical information used in training and inference. 

The feature map and variational layers can take different forms, called \emph{ansätze}, consisting of many possible different quantum gates in different configurations. Such immense freedom raises an important question: how should one choose an architecture for a given problem, and can it be expected to yield a quantum advantage? This question is of major practical importance, and although benchmark results have been shown for very particular datasets \cite{ibm, advantage}, there is little consensus on which ansätze are good choices for machine learning. 

In our work, we use quantum neural networks with orthogonal and compound layers. Although these neural networks roughly match the general VQC construction, they produce well-defined linear algebraic operations, which not only makes them much more interpretable but gives us the ability to analyze their complexity and scalability. Because we understand the actions of these layers precisely, we are able to identify instances for which we can design efficient classical simulators, allowing us to classically train and test the models on real-scale datasets.

A standard feed-forward neural network layer modifies an input vector by first multiplying it by a weight matrix and then applying a non-linearity to the result. Feed-forward neural networks usually use many such layers and learn to predict a target variable by optimizing the weight matrices to minimize a loss function. Enforcing the orthogonality of these weight matrices, as proposed in \cite{jia2019orthogonal}, brings theoretical and practical benefits: it reduces the redundancy in the trained weights and can avoid the age-old problem of vanishing gradients. However, the overhead of typical projection-based methods to enforce orthogonality prevents mainstream adoption.

In \cite{Landman2022quantummethods}, an improved method of constructing orthogonal neural networks using quantum ideas was developed. We describe it below in brief.

\subsection{Data Loaders}
\label{sec:quantum_data_loaders}

In order to perform a machine learning task with a quantum computer, we need to first load classical data into the quantum circuit.

\subsubsection*{Unary data loading circuits}

\begin{figure}[ht]
    \centering
    \includegraphics[width=0.8\textwidth]{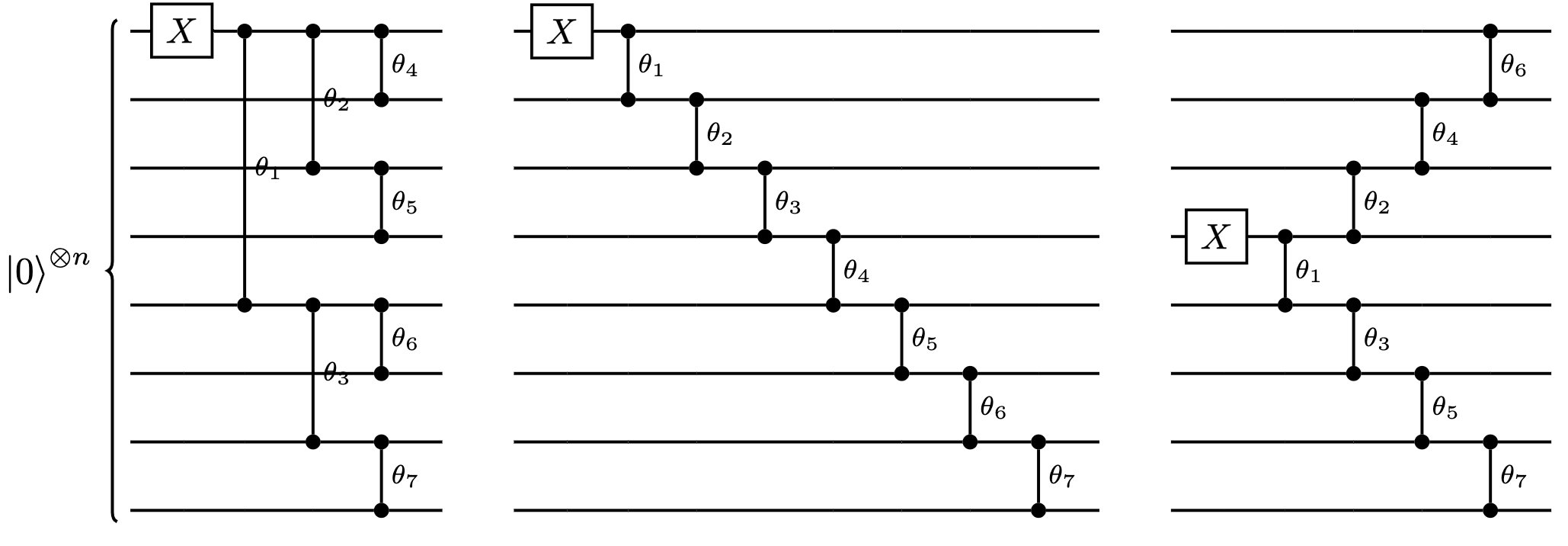}
    \caption{Three possible unary data loaders for $d$-dimensional vectors ($d=8$). From left to right: the parallel, diagonal, and semi-diagonal circuits have respectively a circuit depth of $log(d)$, $d$, and $d/2$. The X gate represents the Pauli X gate, and the vertical lines represent $RBS$ gates with tunable parameters.}
    \label{fig:dataloaders_vector}
\end{figure}

The first way we will load classical data is an example of \emph{amplitude encoding}, which means that we load the (normalised) vector elements as the amplitudes of a quantum state. In \cite{NearestCentroid2021}, three different circuits to load a vector $\bm{x} \in \mathbb{R}^d$ using $d-1$ gates are proposed. The circuits range in depth from $O(log(d))$ to $O(d)$, with varying qubit connectivity (see Fig.\ref{fig:dataloaders_vector}). They use the \emph{unary} amplitude encoding, where a vector $\bm{x} = (x_1,\cdots,x_d)$ is loaded in the quantum state $\ket{\bm{x}} = \frac{1}{\|x\|}\sum_{i=1}^d x_i\ket{e_i}$, where $\ket{e_i}$ is the quantum state with all qubits in $\ket{0}$ except the $i^{th}$ qubit in state $\ket{1}$ (e.g., $\ket{e_3} = \ket{00100000}$). The circuit uses $RBS$ gates: a parameterized two-qubit hamming weight-preserving gate implementing the unitary given by Eq.\ref{RBS}:

\begin{equation} \label{RBS}
RBS(\theta) = \left( \begin{array}{cccc}
1 & 0 & 0 & 0 \\
0 & \cos \theta & \sin \theta & 0 \\
0 & -\sin\theta & \cos\theta & 0 \\
0 & 0 & 0 & 1  \end{array} \right) \ .
\end{equation}

The parameters $\theta_i: i \in \{1,...,d-1\}$ of the $d-1$ RBS gates are classically pre-computed to ensure they encode the correct vector $\ket{\matr{x}}$.

\subsubsection*{$RY$-loading circuits}
We will also use data loading procedures beyond the unary basis.
In particular, for a normalized input vector $\bm{x} \in \mathbb{R}^d$, we use $d$ qubits, where on each of the qubits, we apply an $RY(\theta)$ rotation gate where the angle parameter on the $i^{th}$ qubit is $\theta_i = 2\pi x_i$, according to Eq. \ref{eq:ry}. Multiplication with $2\pi$ allows us to cover the entire range of the $\sin$ and $\cos$ functions. This technique loads the data in the entire $2^d$-dimensional Hilbert space encompassing all the hamming weights from $0$ to $d$. This loading technique has constant depth independent of $d$, and we refer to it as the \emph{RY loading}, whose circuit for $d=8$ is illustrated in Fig. \ref{fig:loaders}.

\begin{equation}
\label{eq:ry}
    RY(\theta)\ket{0} = \cos{\frac{\theta}{2}}\ket{0} + \sin{\frac{\theta}{2}}\ket{1}
\end{equation}

\subsubsection*{$H$-loading circuits}
\begin{figure}[h]
    \centering
    \begin{minipage}{0.45\textwidth}
        \centering
        \includegraphics[width=\textwidth]{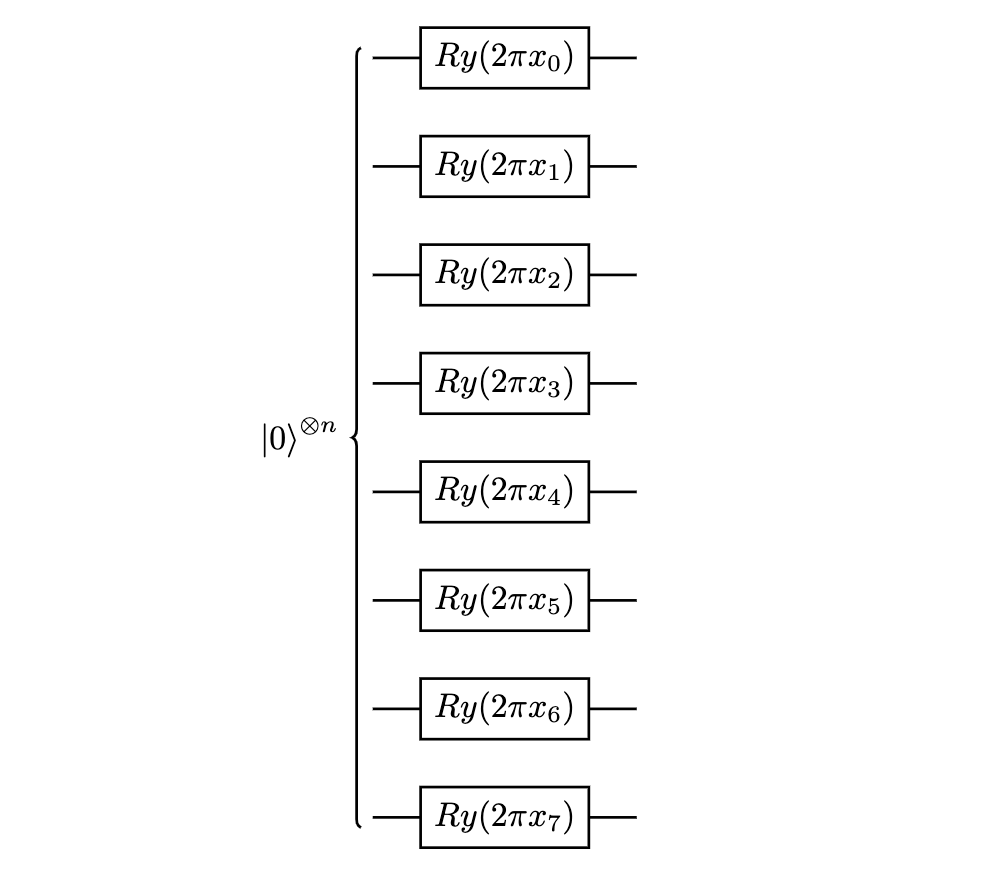}
        \\(a) RY Loader\vspace{0.3cm}
    \end{minipage}
    \hfill
    \begin{minipage}{0.45\textwidth}
        \centering
        \includegraphics[width=0.75\textwidth]{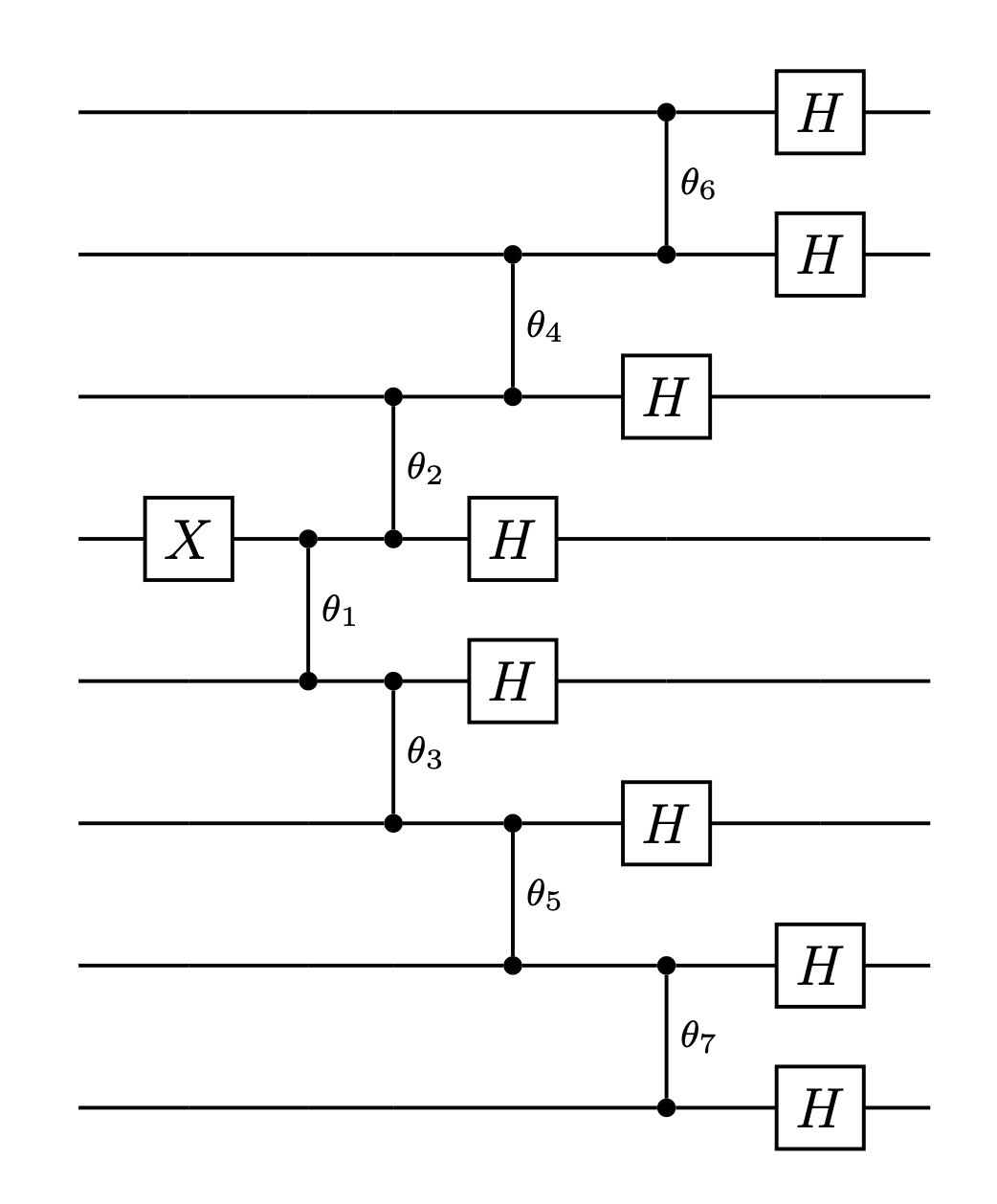}
        \\(b) Hadamard Loader\vspace{0.3cm}
    \end{minipage}
    \caption{Non-Unary Loaders.}
    \label{fig:loaders}
\end{figure}

Lastly, we define a different technique for loading the data in the entire $2^d$-dimensional Hilbert space, which loads the vector in the unary basis and then applies a Hadamard gate on each qubit. This operation applies a Fourier transform on $\mathbb{Z}_2$ and gives us a state encompassing all the hamming weights from $0$ to $d$ at no additional cost to the circuit depth. We call this the $H$-\emph{loading}, whose circuit for $d=8$ is illustrated in Fig. \ref{fig:loaders}. 

\begin{equation}
\label{eq:h}
    H\ket{0} = \frac{\ket{0}+\ket{1}}{\sqrt{2}} \hspace{1cm}
    H\ket{1} = \frac{\ket{0}-\ket{1}}{\sqrt{2}}    
\end{equation}

\subsubsection*{}
The $RY$ and $H$ loading circuits spread the data over all $2^n$ bases to allow future RBS-based neural network layers to utilize an exponential space. In contrast, unary loaders spread the data across the $n$ bases with hamming weight 1, and since RBS gates are hamming-weight-preserving (known as \textit{match gates} \cite{Jozsa_2008_Match}), they cannot change this. Their action when using unary loaders is thus restricted to a much smaller space.

\subsection{Quantum Orthogonal and Compound Layers}
\label{sec:quantum_ortho_layers}

\begin{figure}[h]
    \centering
    \begin{minipage}{0.27\textwidth}
        \centering
        \includegraphics[width=\textwidth]{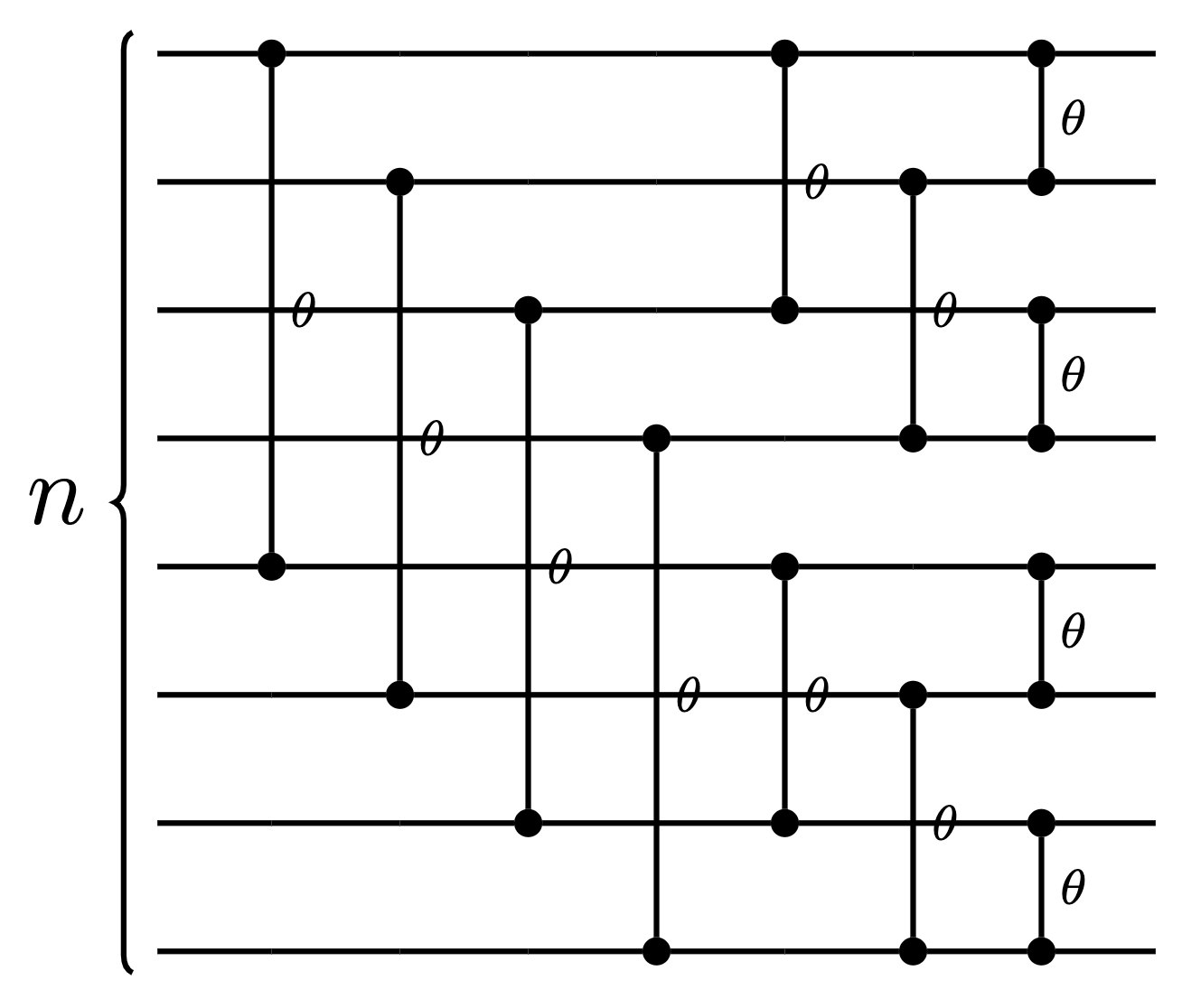}
        (a) Butterfly \vspace{0.3cm} % This adds the label below the image
    \end{minipage}
    \hfill
    \begin{minipage}{0.425\textwidth}
        \centering
        \includegraphics[width=\textwidth]{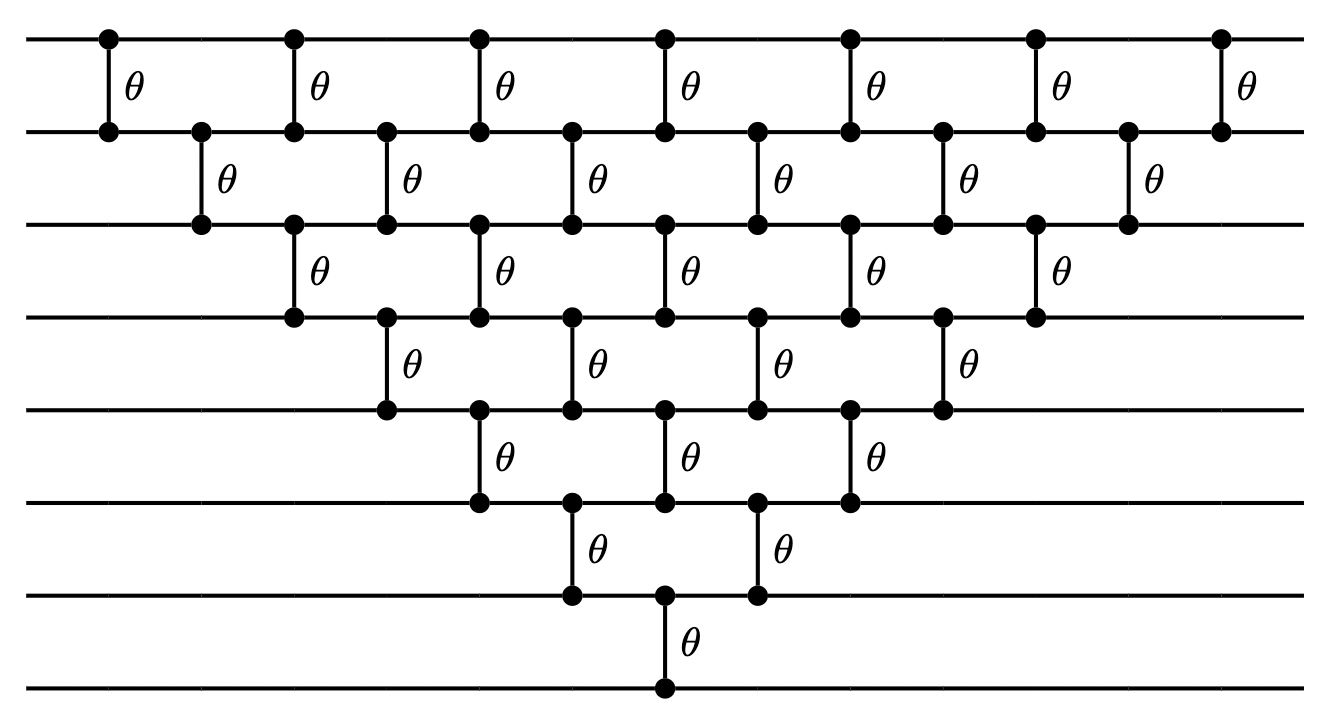}
        (b) Pyramid \vspace{0.3cm} % This adds the label below the image
    \end{minipage}
    \hfill
    \begin{minipage}{0.24\textwidth}
        \centering
        \includegraphics[width=\textwidth]{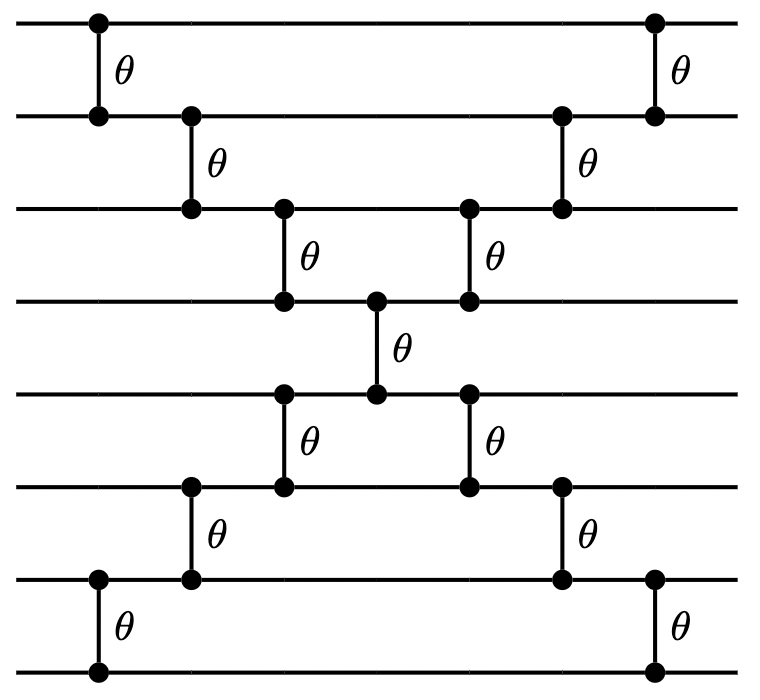}
        (c) X \vspace{0.3cm} % This adds the label below the image
    \end{minipage}
\caption{Parameterized Quantum Circuits for Orthogonal and Compound Layers. Vertical lines represent two-qubit RBS gates, parameterized with independent angles $\theta$, which are shown as the same in the figures for simplicity.}
\label{fig:quantum_circuits}
\end{figure}

Quantum orthogonal layers consist of a unary data loader plus a parametrized quantum circuit made of $RBS$ gates, while quantum compound layers consist of a general data loader plus a parametrized quantum circuit made of $RBS$ gates. 

$RBS$ gates and circuits preserve the hamming weight of the input state, and thus if we use a unary data loader, then the output of the layer will be another vector in unary amplitude encoding. Similarly, if the loaded quantum state is a superposition of only basis states of hamming weight $k$, so is the output state.
More generally, we can think of such hamming-weight preserving circuits with $n$ qubits as block-diagonal unitaries that act separately on $n+1$ subspaces, where the $k^{th}$ subspace is defined by all computational basis states with hamming weight equal to $k$. The dimension of these subspaces is equal to $n \choose k$.
The first block of this unitary is an ${n \times n}$ orthogonal matrix, such that when a vector is loaded in the unary basis, this circuit simply performs orthogonal matrix multiplication.
In general, the $k$-th block of this unitary applies a \emph{compound matrix of order k} of the $n \times n$ unary matrix. The dimension of this $k$-th order compound matrix is ${n \choose k} \times {n \choose k}$. We refer to the layers that use bases beyond the unary as \emph{compound layers.}

There exist many possibilities for building a parametrized quantum circuit made of $RBS$ gates which can be used in a quantum orthogonal or compound layer, each with different properties. 

The \textbf{Pyramid circuit} (Fig.\ref{fig:quantum_circuits}), proposed in \cite{Landman2022quantummethods}, is a parameterised quantum circuit composed of exactly $n(n-1)/2$ $RBS$ gates. This circuit requires only adjacent qubit connectivity, which makes it suitable for most superconducting qubit hardware. In addition, when restricted to the unary basis, the pyramid circuit expresses exactly the Special Orthogonal Group, i.e. orthogonal matrices with the determinant equal to $+1$. To allow this circuit to express the entire orthogonal group, we can add a final $Z$ gate on the last qubit. This allows us to express orthogonal matrices with a $-1$ determinant as well. The pyramid circuit is, therefore, very general and covers all the possible orthogonal matrices of size $n \times n$.

The \textbf{$\mathbf{X}$ circuit} (Fig.\ref{fig:quantum_circuits}), introduced in \cite{cherrat2022quantum}, uses just $O(n)$ gates and has nearest-neighbor connectivity. Due to reduced depth and gate complexity, it accumulates less hardware noise.

The \textbf{Butterfly circuit} (Fig.\ref{fig:quantum_circuits})
is inspired by the classical fast Fourier transform algorithm, and uses $O(n log(n))$ gates. It was also introduced in \cite{cherrat2022quantum}, and despite having reduced expressivity compared to the Pyramid circuit, it often performs just as well.

In \cite{Landman2022quantummethods}, a method is proposed to train orthogonal layers for the unary basis by computing the gradient of each parameter $\theta_i$ using backpropagation. This backpropagation method for the pyramid circuit (which is the same for any circuit with $RBS$ gates) takes time $O(n^2)$, corresponding to the number of gates, and provides a polynomial improvement in runtime compared to the previously known orthogonal neural network training algorithms which relied on an $O(n^3)$ SVD operation \cite{jia2019orthogonal}. Since the runtime corresponds to the number of gates, it is lower for the butterfly and $X$ circuits. See Table \ref{table:circuit_comparison} for full details on the comparison between the three types of circuits.
For the compound layers, we need to consider the entire $2^n \times 2^n$ space and thus train an exponential size weight matrix, which takes exponential time on a classical computer. In principle, a compound layer can also be trained using the parameter shift rule for quantum circuits, which can be more efficient since the number of parameters is polynomial in the input size, though noise in current quantum hardware makes this impractical for the time being. 

\begin{table}[h]
\caption{Comparison of different parameterized quantum circuits for orthogonal and compound layers with $n$ qubits.}\label{table:circuit_comparison}
\begin{tabular}{@{}p{0.15\textwidth}p{0.25\textwidth}p{0.15\textwidth}p{0.15\textwidth}@{}}
\toprule
Circuit & Hardware Connectivity & Depth & \# Gates \\
\midrule
Pyramid & Nearest-Neighbor & $2n-3$ & $\frac{n(n-1)}{2}$ \\
X & Nearest-Neighbor & $n-1$ & $2n-3$ \\
Butterfly & All-to-all & $log(n)$ & $\frac{n}{2} log(n)$ \\
\botrule
\end{tabular}
\end{table}

\subsection{Expectation-per-Subspace Compound Layer}
\label{sec:exp_layer}

\begin{figure}[h]
    \centering
    \includegraphics[width=\textwidth]{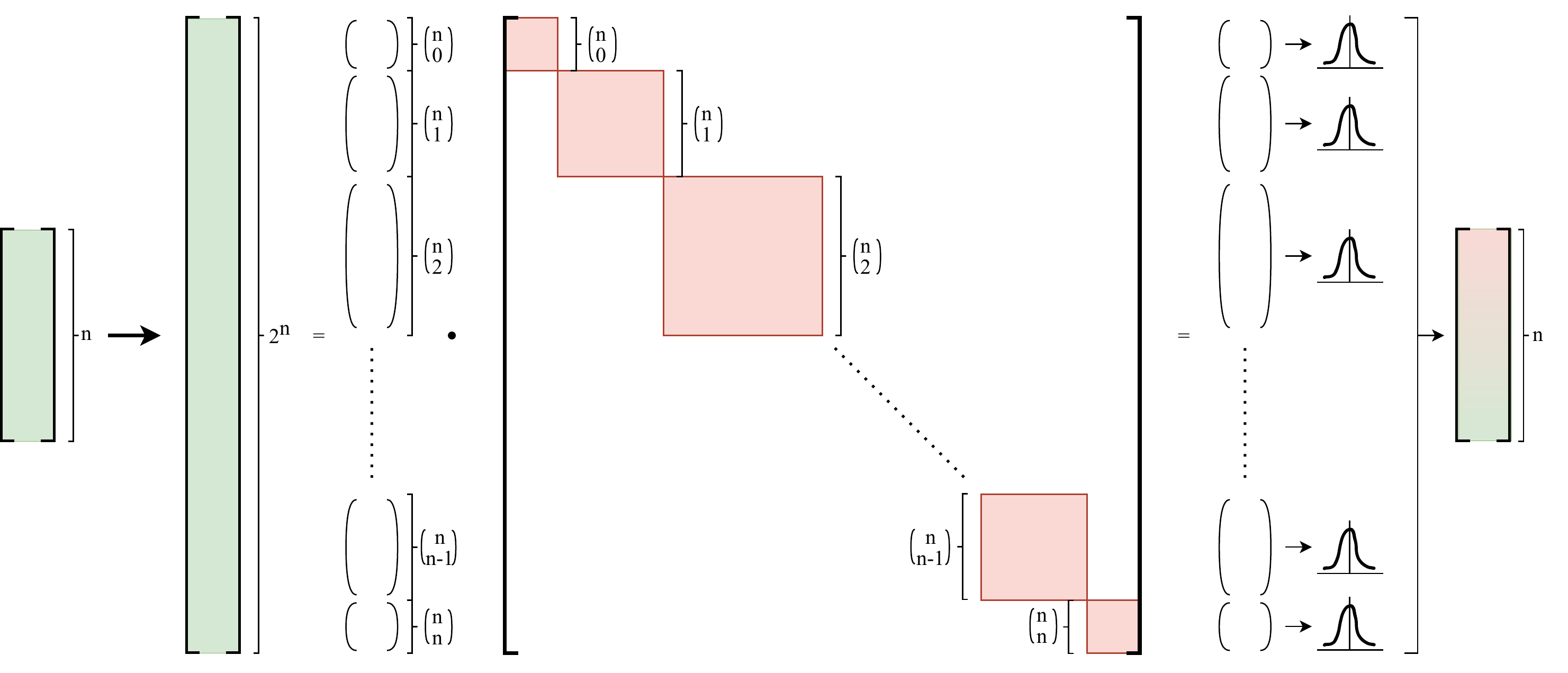}
    \caption{Expectation-per-subspace Compound Layer. In the final step, we combine the $n \choose 0$ and the $n \choose n$ subspaces and calculate their overall expectation.}
    \label{fig:expectation_layer}
\end{figure}

We describe here a compound layer that we call the Expectation-per-subspace compound layer. This layer involves loading the input vector using a non-unary basis which could be done either via the $RY$-loading or the $H$-loading circuit as previously defined. Then, we apply a parameterized quantum circuit with RBS gates, e.g. a pyramid circuit, which performs the compound matrix operation on all the fixed hamming weight subspaces. More precisely, we can think of the operation as performing the matrix-vector multiplication of an ${n \choose k} \times {n \choose k}$ matrix with an ${n \choose k}$-dimensional vector for each hamming weight $k$ from $0$ to $n$. Note that for $0$ and $n$, the dimension is $1$, and hence the unitary acts as identity.

If we look at the output quantum state, it defines a distribution over a domain of size $2^n$. Given the exponential size of the distribution, it is not advisable to try and train the entire distribution, since that would take exponential time. However, one can still try to use a loss function that contains some information about the distribution. For example, one can use the expectation of the distribution, which is what normally happens in variational quantum algorithms where one approximates this expectation by using a number of measurement outcomes. Given the fact that our unitary is block-diagonal, one can try to define a more complex loss function that contains more information about the distribution. In particular, one can split the domain of the distribution into $n+1$ subdomains, one for each subspace, and then train on all these expectations.

This is what we do in the Expectation-per-subspace compound layer, where for each $k$ from $0$ to $n$, we take the outputs corresponding to the hamming weight $k$ strings and sort them. Now, for each $k$, we assign values which are equally spaced between two bounds $a$ and $b$ (which are $0$ and $10$, in our models) to the $n \choose k$ strings. We normalize the outputs using the $L1$-norm to correspond to a probability distribution over the $n \choose k$ values between $a$ and $b$, and then we calculate the expectation value for that hamming weight. This gives us a set of $n+1$ values corresponding to each hamming weight. Since for hamming weight $0$ and $n$ the dimension of the subspace is 1 (the all-zero and all-one strings), we combine them and calculate the expectation for these two together and make the layer have $n$ outputs. The entire operation is illustrated in Fig. \ref{fig:expectation_layer}.

While these compound layers do increase classical simulation complexity, they do not increase quantum complexity. The advent of better quantum hardware will allow us to test larger compound layers that explore much larger portions of the Hilbert space.

\section{QNN Results with Classical Simulation}\label{sec:simulation_results}

In this second study, we focus on the problem of credit-default prediction associated with credit applications from Small and Medium Enterprises (SMEs). We report the results of our neural network architectures on this use case and on public datasets.

\subsection{Use Case Introduction: Credit Risk Prediction}\label{sec:sme_credit_usecase}

The credit operation is one of the largest and most important operations in a retail banking institution. Throughout the credit journey (life-cycle) of a customer within the bank, several different models are used at different points of the journey and for different purposes, such as the determination of interest rates, offering of different products, etc. 

The credit granting model is a particularly important one since it determines whether or not a credit relationship will be established. It is also particularly challenging in the case of SMEs, where the relationship with the bank often starts only when the SME submits an application for credit, so very little data is available. 

Given these challenges, we propose the use of quantum techniques aiming at improving the predictive performance of the credit granting model.

The credit granting decision may be seen as a binary classification problem, in which the objective is to predict if the SME will default on credit. More specifically, we are interested in calculating the so-called probability of default (PD), which is given by $P(\hat{y}=1|\bm{x})$. The PD information is used internally for other pipelines primarily concerned with the determination of credit ratings for the SMEs (though in this study, we focus solely on the PD model), so the PD distribution is the main output of interest from the model. For this reason, we do not threshold the probability outputs from the model --- thus, we use threshold-independent classification metrics to evaluate its predictive performance. Namely, the main Key Performance Indicator (KPI) that we use is the Gini score, constructed from the Area Under the Curve (AUC) of the ROC (Receiver Operating Characteristic) curve as $\text{Gini} = 2 \times \text{AUC} - 1$. The Gini score is easily interpreted by the business team and allows for a holistic estimation of the model's impact.

In this study, we chose to focus on the development of an "internal model" of credit default, which only uses features collected by Itaú, without considering any external information (from credit bureaus, for instance). The dataset used consists of $\approx 141,500$ observations, each one represented by 32 features: 31 numerical and 1 categorical. Each observation represents a given SME customer in a specified reference month, whose observed target indicates its default behavior, and whose features consist of internal information about the company. The data was anonymized, standardized, and split into training and test sets based on the time period: the training set consists of $\approx 74,700$ observations covering 12 months of data, while in the test set, we have $\approx 66,800$ observations covering the subsequent 8 months.

\subsection{Neural Network Architectures for Credit Risk}\label{sec:orthonn_usecase_architecture}

To compare the performance of the orthogonal and compound layers to the classical baseline, we designed three neural network architectures. Each architecture had three layers: an \textit{encoding layer}, an \textit{experimental layer}, and a \textit{classification head}. The encoding layer was a standard linear layer of size $32 \times 8$ followed by a $\tanh$ activation. Its purpose is to bring the dimension of the features down to 8, which is a reasonable simulation size for both proposed quantum layers. The second layer was the experimental layer of size $8 \times 8$ (described below). Finally, the third layer, the classification head, was a linear layer of size $8 \times 2$ followed by softmax to predict the probabilities.

\begin{figure}[H]
    \centering
    \includegraphics[width=.7\linewidth]{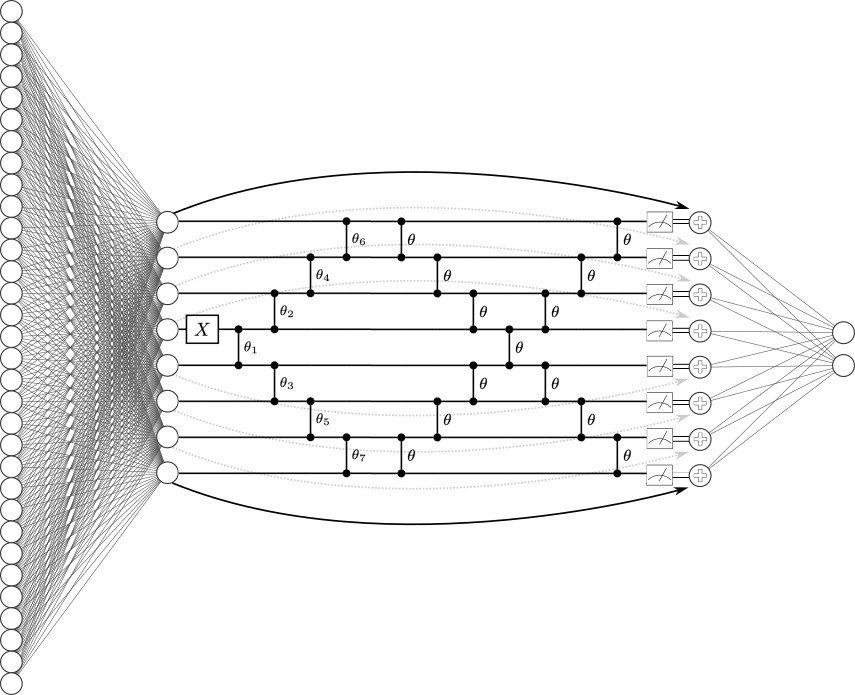}
    \caption{Architecture of the OrthoResNN model.}
    \label{fig:ortho_res_nn_arc}
\end{figure}

The first quantum neural network architecture, named \textbf{OrthoResNN}, uses an $8 \times 8$ orthogonal experimental layer implemented with a semi-diagonal loader and $X$ circuit. Note that the final output of the layer is provided by measurements. We add a skip connection by adding the input of the orthogonal layer to the output \footnote{The model cannot learn to "ignore" the orthogonal layer via the skip connection, because the orthogonal layer cannot change the magnitude of the input.}. The layer is followed by a $\tanh$ activation function. This architecture is illustrated in Fig. \ref{fig:ortho_res_nn_arc}.

\begin{figure}[H]
    \centering
    \includegraphics[width=.65\linewidth]{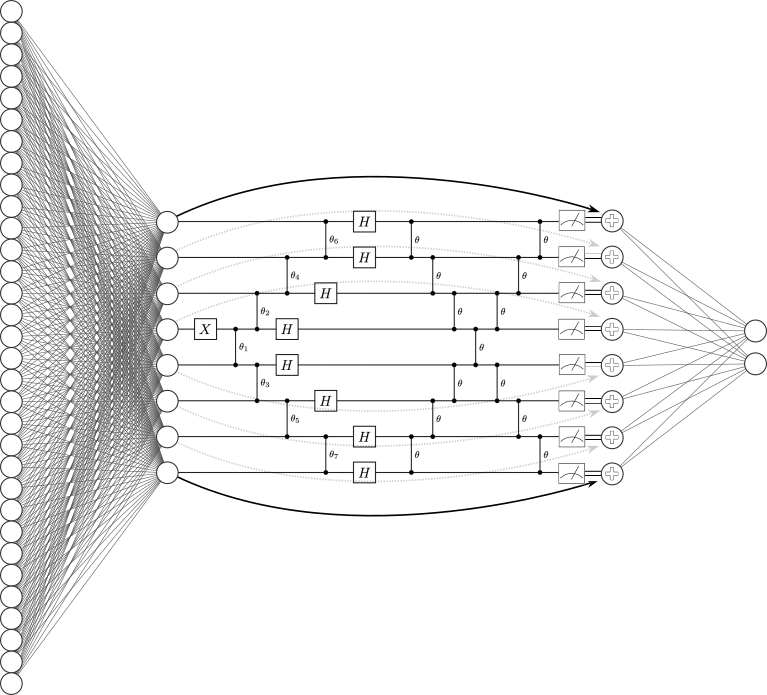}
    \caption{Architecture of the ExpResNN model using the $H$-loader.}
    \label{fig:exp_res_nn_hadamard}
\end{figure}

\label{sec:expnn}
Our second architecture, \textbf{ExpResNN}, replaces the experimental layer with an $8 \times 8$ Expectation-per-subspace compound layer. We use the the $H$-loader to encode our data. The layer is again followed by a $\tanh$ activation function. Fig. \ref{fig:exp_res_nn_hadamard} illustrates the ExpResNN architecture.

And finally, the classical architecture, \textbf{ResNN}, used an $8 \times 8$ linear residual layer followed by $\tanh$.

\subsection{Methods and Training}

The training of the networks was performed using the \texttt{JAX} package by Google. We train our models for $500$ epochs. To identify suitable hyperparameters, we performed a search over learning rate, learning-rate-halving points, and batch size. The hyperparameter search was performed with the \texttt{ray-tune} library.

The dataset contains a large number of missing values, which motivated the experimentation of different imputation techniques such as zero-filling, round-robin imputation (implemented in the Python package \texttt{scikit-learn}), and MICE \cite{mice}. The best results were achieved with round-robin imputation using \texttt{scikit-learn}'s \texttt{IterativeImputer} with Bayesian ridge regression. This was the pre-processing employed in all the results on the SME dataset. 

\subsection{Results}\label{sec:sme_credit_usecase}

In our experimental setup, we consider the fully connected residual layer (ResNN) as the classical benchmark. We performed the same experiment with an orthogonal layer using the semi-diagonal loader and the X circuit (OrthoResNN). Finally, we tried the expectation-per-subspace compound layer with the Hadamard loader and X circuit (ExpResNN). While the performance of the OrthoResNN and ExpResNN remained nearly the same as the FNN layer, these new layers learn the angles of $2n$ RBS gates instead of $n^2$ elements of the weight matrix, dramatically reducing the number of parameters needed. The results are shown in Table \ref{table:fnn_sim_results}.

\begin{table}[h]
\caption{Comparison between different architectures.}\label{table:fnn_sim_results}
\begin{tabular}{@{}p{0.155\textwidth}p{0.25\textwidth}p{0.15\textwidth}@{}}
\toprule
Model & Trainable parameters & Test Gini \\
\midrule
ResNN & 64 & 54.20 \% \\
OrthoResNN & 13 & 54.29\% \\
ExpResNN & 13 & 53.95\% \\
\botrule
\end{tabular}
\end{table}

The results show that quantum orthogonal and compound layers can preserve the performance of fully-connected layers on this dataset while using a fraction of the trainable parameters. We note that the ExpResNN did not show advantages over the OrthoResFNN. 

\subsection{Further Benchmarks}
\label{sec:further_benchmarks_nn}

We compared the orthogonal and linear layers on public classification datasets from OpenML as we did for the Random Forests (Sec. \ref{sec:further_benchmarks_rf}). Again, we used a three-layer architecture: a linear encoding layer to $16$ dimensions with GeLU activation, a $16 \times 16$ experiment layer with tanh activation, and a binary classification head. These architectures did not have residual connections.

We compared two models: \textbf{OrthoFNN} and \textbf{FNN}. OrthoFNN used a $16 \times 16$ pyramid circuit as the experiment layer, for a total of 120 trainable parameters. FNN used a feed-forward linear layer with 256 trainable parameters.

For all the datasets and models, we use a batch size of $128$ and a learning rate of $10^{-4}$. Each is trained for $500$ epochs and evaluated with the ROC-AUC metric. The results are summarised in Table \ref{table:NN_comparison}.

\begin{table}[h]
\caption{Comparison of OrthoFNN and FNN for different datasets. The results are reported via the ROC-AUC metric.}\label{table:NN_comparison}
\begin{tabular}{@{}p{0.25\textwidth}p{0.15\textwidth}p{0.15\textwidth}@{}}
\toprule
Dataset & FNN & OrthoFNN \\
\midrule
madelon & 0.624 & \textbf{0.636} \\
credit-default & \textbf{0.647} & 0.634 \\
house-pricing & 0.780 & \textbf{0.781} \\
jannis & 0.791 & \textbf{0.797} \\
eye movements & \textbf{0.580} & 0.578 \\
bank-marketing & 0.847 & \textbf{0.850} \\
wine & 0.553 & \textbf{0.640} \\
california & 0.634 & \textbf{0.668} \\
\botrule
\end{tabular}
\end{table}

\section{QNN Results on Quantum Hardware}

\subsection{Implementation of Quantum Circuits}

Using a classical computer, inference using an orthogonal layer takes time $O(n^2)$, while for a general compound layer, this time is exponential in $n$. Using a quantum computer, inference with an orthogonal or compound layer uses a quantum circuit that has depth $O(n)$ (Pyramid or X) or $O(\log(n))$ (Butterfly), and $O(n^2)$ gates. Therefore, one may find a further advantage if the inference is performed on a quantum computer. This motivated us to test the inference step for classically trained OrthoResNN and ExpFNN models (ExpResNN from the classical experiments without a residual connection) on currently available quantum hardware.

\begin{figure}[!h]
    \centering
    \includegraphics[width=0.61\textwidth]{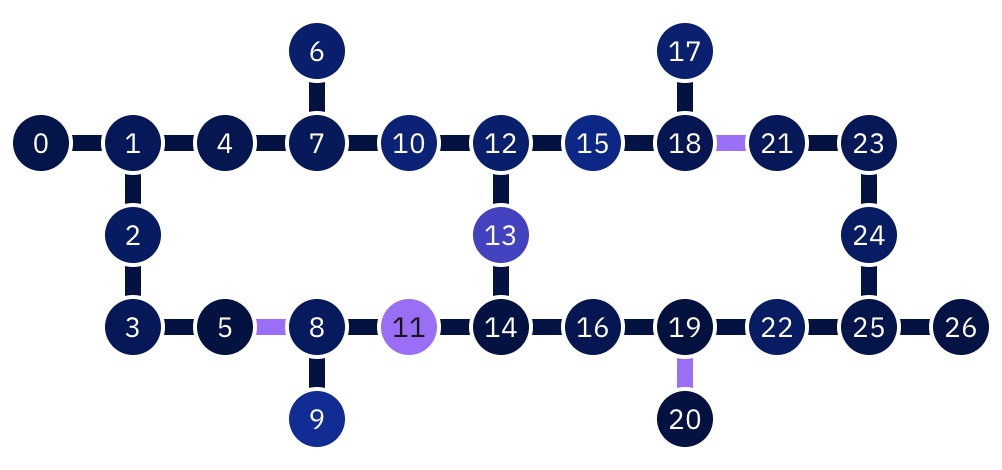}
    \caption{Topology graph of the 27-qubit \emph{ibm\_hanoi} machine used to perform our hardware experiments. The colors in the qubits indicate readout assignment error; and in the connections the CNOT error --- dark blue is low, purple is high.}
    \label{fig:ibm_hanoi}
\end{figure}

The data loader and orthogonal/compound layer circuits employed in our model architectures are NISQ-friendly and particularly suitable for superconducting qubits, with low depth and nearest-neighbors qubit connectivity. Thus, we chose to use IBM's 27-qubit machine ibm\_hanoi (see Fig.\ref{fig:ibm_hanoi}).

To perform inference on \emph{ibm\_hanoi}, we used the semi-diagonal data loader and X circuit to implement the OrthoResNN model; and the Hadamard loader and X circuit for the ExpFNN model --- the same architectures described in Sec. \ref{sec:simulation_results}. Both neural networks were trained classically, and the trained parameters were used to construct our quantum circuits for inference.

Given the large size of the test dataset ($66,750$ data points), we decided to perform inference using the trained models on a small test subsample of $300$ test points, corresponding to the maximum number of circuits we could send in one job to the IBM machine. After testing different subsamples with the OrthoResNN model\footnote{We made sure to pick a representative subsample of observations, for which we would have neither a heavily under nor overestimated Gini score compared to the one for the full test set. By selecting 50 different subsamples of randomly chosen 300 observations and performing the classical inference, we found Gini scores between 45.9\% and 64.3\%, with an average of 54.44\%. This further supports the fact that the subsample that we chose correctly yields the (approximate) expected value for this Gini score distribution, which then yields an unbiased subsample Gini score. The same procedure was performed for the ExpFNN experiment.},
we selected one for which we achieved a subsample test Gini score of 54.19\% using a noiseless simulator (blue ROC curve in Figure \ref{fig:roc_curves}). The same was done for the ExpFNN experiment, yielding a subsample test Gini of 53.90\% with the noiseless simulator. These values were taken as the best possible Gini scores if the inference was performed on noiseless quantum hardware, which could then be compared with the values actually achieved with ibm\_hanoi.
% The subsamples were chosen so that the Gini scores matched that of the full dataset.

The circuits were then run on the quantum processor. Due to its limited Hilbert space of size $n$, the OrthoResNN has a natural error-rejection procedure: any measurements outside of the unary basis can be disregarded as errors. As a result, the inference yielded a Gini score of $50.19\%$, as shown in the orange ROC curve in Fig. \ref{fig:roc_curves}. The achieved Gini was not too far from the noise-free simulation result (54.19\%), but there was clearly room for improvement in order to close the 4 pp difference. We also attempted inference with the more complex ExpFNN, which yielded a Gini score of 40.20\%, much farther from the noiseless simulation Gini of 53.90\%. Since the ExpFNN uses the entire $2^n$-dimensional Hilbert space, it is more prone to errors due to noise, as the error-rejection procedure used for the OrthoResNN cannot be employed.

\subsection{Improving the Hardware Results with Error Mitigation Techniques}

Error mitigation and error suppression techniques undoubtedly play a very important role in NISQ-era quantum computing. While these techniques alone may not be sufficient to fully overcome the imperfections of current quantum systems, they can push the practical limits of what can be achieved. As a next step, for the OrthoResNN model, we experimented with various error mitigation and suppression approaches, going beyond the simple hamming-weight postselection procedure, in an attempt to close the gap of 4 pp between the Gini score from the noisy simulation and the one from hardware execution.

The first approach that we tried was a correlated readout mitigator. This is a purely classical post-processing technique which demands the construction of a calibration circuit for each one of the possible $2^N$ states of the full $N$ qubits Hilbert space. The calibration circuits' execution (simulated using ibm\_hanoi's backend information, in our case) yields a $2^N \times 2^N$ assignment matrix, which is used to understand how errors might occur during readout. One can see that this method rapidly becomes intractable as the number of qubits $N$ increases. In our case, for $N=8$, the Gini score improved to 50.24\%, a small improvement of only 0.05 pp.

Thus, in order to investigate the effect of more robust error suppression and mitigation techniques in our results, we moved on to a new round of hardware experiments, performing the inference by executing the exact same OrthoResNN circuits via the Qiskit Runtime \cite{Qiskit} service using the Sampler primitive, which allows one to use circuit optimization as well as error suppression and mitigation techniques, as detailed below.

Firstly, we used circuit optimization at the point of circuit transpilation and compilation by setting the optimization\_level parameter to the highest possible value, 3. This performs the following circuit optimization routines: Layout selection and routing (VF2 layout pass and SABRE layout search heuristics \cite{sabre}; 1 qubit gate optimization (chains of single-qubit u1, u2, u3 gates are combined into a single gate); commutative cancellation (cancelling of redundant self-adjoint gates); 2 qubit KAK optimization (decomposition of 2-qubit unitaries into the minimal number of uses of 2-qubit basis gates).

Secondly, we used the Dynamical Decoupling error suppression technique  \cite{DD_seminal, DD_survey}. This technique works as a pulse schedule by inserting a DD pulse sequence into periods of time in which qubits are idle. The DD pulses effectively behave as an identity gate, thus not altering the logical action of the circuit, but having the effect of mitigating decoherence in the idle periods, reducing the impact of errors.

Thirdly, we used the M3 (Matrix-free Measurement Mitigation) error mitigation technique \cite{M3} by setting the Sampler resilience\_level parameter to 1 (the only option available for the Sampler primitive). This provides mitigated quasi-probability distributions after the measurement. M3 works in a reduced subspace defined by the noisy input bitstrings supposed to be corrected, which is often much smaller than the full $N$ qubits Hilbert space. For this reason, this method is much more efficient than the matrix-based readout mitigator technique mentioned above. M3 provides a matrix-free preconditioned iterative solution method, which removes the need to construct the full reduced assignment matrix but rather computes individual matrix elements, which uses orders of magnitude less memory than direct factorization.

\begin{figure}[!ht]
    \centering
    \includegraphics[width=0.8\textwidth]{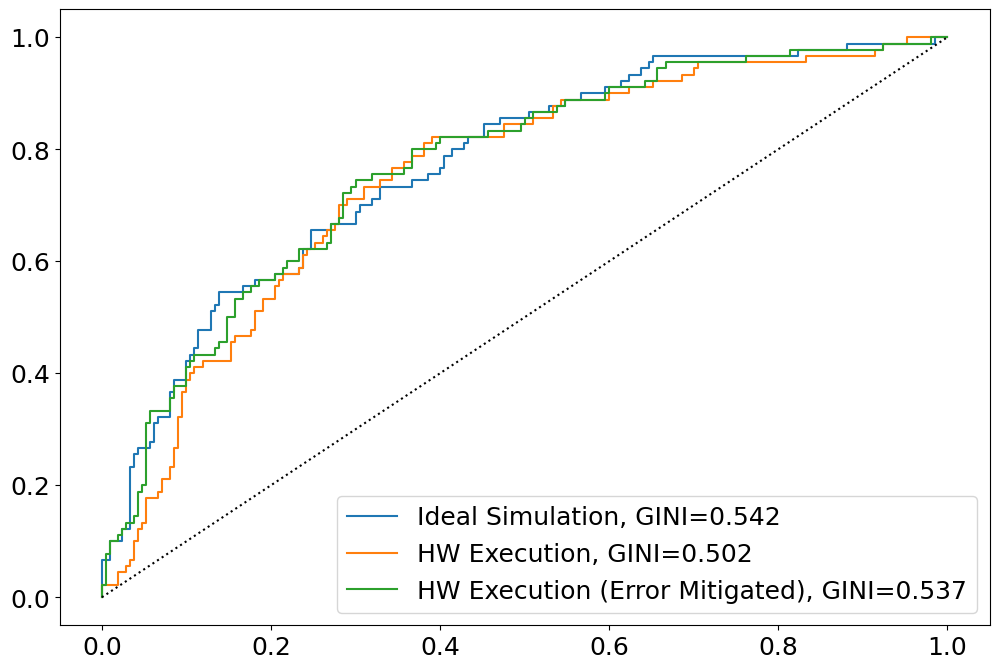}
    \caption{ROC Curves with Gini score for the ideal simulation, hardware execution, and the error-mitigated hardware execution.}
    \label{fig:roc_curves}
\end{figure}

By employing these three techniques, we were able to achieve a Gini score of 53.68\% for the OrthoResNN (Fig. \ref{fig:roc_curves}). This is a 3.49 pp improvement from the initial 50.19\% Gini of the unmitigated run, falling only 0.53 pp behind the ideal noiseless execution (54.21\% Gini)! This remarkable result underscores the NISQ-friendliness of the orthogonal layer and highlights the importance of error suppression and mitigation techniques in the NISQ era.

It is important to note that circuit optimization, error suppression, and mitigation techniques typically result in some classical/quantum pre/post-processing overhead to the overall circuit runtime. Some of these techniques are based on heuristics and/or do not have efficient scaling at larger circuit sizes. It is important to balance the desired levels of optimization and resilience with the required time for the full execution, especially as the circuit sizes increase.

\section{Conclusion}

In this work, we have explored the potential of quantum machine learning methods in improving forecasting in finance, with a focus on two specific use cases within the Itaú business: churn prediction and credit risk assessment. Our results demonstrate that the proposed algorithms, which leverage quantum ideas, can effectively enhance the performance of Random Forest and neural network models, achieving better accuracy and training with fewer parameters.

In the present day, quantum hardware is not powerful enough to provide real improvements or conclusive large-scale benchmarks. Performance enhancements can be achieved today by turning these quantum ideas into classical ML solutions run on GPUs. However, with the advent of better quantum hardware, we expect these methods to run faster and produce even better results when run on quantum computers.

The general nature of the proposed methods makes them applicable to other use cases in finance and beyond, although they must be tuned to specific datasets and tasks. We hope this work inspires confidence that QML research holds promise both for today as well as for the coming era of scaled, fault-tolerant quantum hardware.

\backmatter

\bmhead{Disclaimer}
This paper is a research collaboration between Itaú Unibanco and QC Ware.
Any opinions, findings, conclusions or recommendations expressed in this material are those of the authors and do not necessarily reflect the views of Itaú Unibanco.
This paper is not and does not constitute or intend to constitute investment advice or any investment service. 
It is not and should not be deemed to be an offer to purchase or sell, or a solicitation of an offer to purchase or sell, or a recommendation to purchase or sell any securities or other financial instruments.
Moreover, all data used in this study is compliant with the Brazilian General Data Protection Law.

\section*{Declarations}

\subsection*{Ethical Approval}
Not applicable as this study did not involve any human or animal subjects.

\subsection*{Competing interests}
S.T., S.K., N.M., and I.K. are employed by QC Ware, while A.F. and S.B. are employed by Itaú, both of which may stand to benefit from the results of this research. The authors declare no other competing interests.

\subsection*{Authors' contributions}
S.T., S.K. N.M., and A.F. wrote the manuscript body and performed the classical experiments. N.M., S.K., and A.F. performed the quantum hardware experiments. All authors reviewed the manuscript.

\subsection*{Funding}
This research did not receive any specific grant from funding agencies in the public, commercial, or not-for-profit sectors.

\subsection*{Availability of data and materials}
The primary datasets and model used in this study are proprietary to Itau and QC Ware and are not publicly available. Additional benchmark datasets are available via openML.

\bibliography{sn-bibliography}

\end{document}